\documentclass[preprint2]{aastex6}
\pdfoutput=1

\usepackage{tabularx}
\usepackage{amsmath}
\usepackage{booktabs}
\usepackage{natbib}

\providecommand{\e}[1]{\ensuremath{\times 10^{#1}}}

\makeatletter
\protected\def\OID#1{\@ifundefined{OID#1}{\GenericError{}{OID#1 not defined!}{}{}}{\csname OID#1\endcsname}}
\def\newOID#1#2{\expandafter\def\csname OID#1\endcsname{#2}}
\makeatother

\newOID{1342263459}{Cha~H$\alpha$~1}
\newOID{1342263489}{ESO~H$\alpha$~559}
\newOID{1342263492}{Hn~13}
\newOID{1342263516}{FR~Tau}
\newOID{1342263934}{J04393364}
\newOID{1342264239}{GM~Tau}
\newOID{1342264240}{CFHT-4}
\newOID{1342264241}{FU~Tau~A}
\newOID{1342265469}{CFHT-20}
\newOID{1342265470}{J04381486}
\newOID{1342268646}{CIDA-1}

\newcommand{\ChaI}{Chamaeleon~I}
\newcommand{\Rout} {R_\mathrm{out}}
\newcommand{\Rin} {R_\mathrm{in}}

\newcommand{\Mdust} {M_\mathrm{dust}}
\newcommand{\Mstar} {M_\mathrm{*}}
\newcommand{\Msun} {M_\odot}
\newcommand{\Tave} {\textless T_{\rm d}\textgreater}
\newcommand{\Td} {T_{\rm dust}}
\newcommand{\Tg} {T_{\rm gas}}
\newcommand{\Teff} {T_{\rm eff}}
\newcommand{\Lbol}{L_{\rm bol}}
\newcommand{\Lsun}{L_{\odot}}
\newcommand{\Lsol}{L_{\odot}}
\newcommand{\TvL}{\Tave - \Lbol}
\newcommand{\TvsL}{\Tave - \Lbol}
\newcommand{\OI}{[OI] 63\,\micron{}}
\newcommand{\HSO}{\textit{Herschel Space Observatory} }
\newcommand{\Herschel}{\textit{Herschel} }
\newcommand{\Hubble}{\textit{Hubble} }
\shorttitle{Small Disks around Stellar-boundary Objects}
\shortauthors{Hendler et al.}

\begin{document}

\title{Hints for Small Disks around Very Low-Mass Stars and Brown Dwarfs\thanks{{\it Herschel} is an ESA space observatory with science instruments provided by European-led Principal Investigator consortia and with important participation from NASA.}}

\author{
    Nathanial P. Hendler,
    Gijs D. Mulders\altaffilmark{1},
    Ilaria Pascucci\altaffilmark{1}
}
\affil{Lunar and Planetary Laboratory, The University of Arizona, Tucson, AZ 85721, USA}
\altaffiltext{1}{Earths in Other Solar Systems Team, NASA Nexus for Exoplanet System Science.}
\and

\author{
    Aaron Greenwood,
    Inga Kamp
}
\affil{Kapteyn Astronomical Institute, University of Groningen, Postbus 800, 9700 AV Groningen, The Netherlands}
\and

\author{
    Thomas Henning
}
\affil{Max Planck Institute for Astronomy, Königstuhl 17, D-69117 Heidelberg, Germany}
\and

\author{
    Fran\c{c}ois M\'enard
}
\affil{Univ. Grenoble Alpes, CNRS, IPAG, F-38000 Grenoble, France}
\and

\author{
    William R.~F. Dent
}
\affil{Atacama Large Millimeter/submillimeter Array (ALMA) Santiago Central Offices, Alonso de C\'ordova 3107, Vitacura, Casilla 763 0355, Santiago, Chile Department of Engineerin}
\and

\author{
    Neal J. Evans II
}
\affil{Department of Astronomy, The University of Texas at Austin, Austin, TX 78712, USA}

\email{equant@lpl.arizona.edu}

\begin{abstract}
The properties of disks around brown dwarfs and very-low mass stars (hereafter VLMOs) provide important boundary conditions
on the process of planet formation and inform us about the numbers and masses
of planets than can form in this regime.  We use the \HSO PACS spectrometer to
measure the continuum and \OI{} line emission towards 11 VLMOs with known disks
in the Taurus and \ChaI\ star-forming regions.  We fit radiative transfer
models to the spectral energy distributions of these sources.  Additionally, we
carry out a grid of radiative transfer models run in a regime that connects the
luminosity of our sources with brighter T~Tauri stars.  We find VLMO disks with
sizes [1.3--78] au, smaller than typical T~Tauri disks, fit well the spectral
energy distributions assuming disk geometry and dust properties are
stellar-mass independent.  Reducing the disk size increases the disk
temperature and we show that VLMOs do not follow previously derived disk
temperature-stellar luminosity relationships if the disk outer radius scales
with stellar mass.  Only 2 out of 11 sources are detected in [OI] despite a
better sensitivity than was achieved for T Tauri stars, suggesting that VLMO
disks are underluminous.  Using thermochemical models we show that smaller
disks can lead to the unexpected \OI{} non-detections in our sample.  The disk
outer radius is an important factor in determining the gas and dust
observables. Hence, spatially resolved observations with ALMA -- to establish
if and how disk radii scale with stellar mass -- should be pursued further.

\end{abstract}


\section{Introduction}

Brown dwarfs are sub-stellar objects that represent the low-mass end of the
star formation process.  Similar to their stellar counterparts, T~Tauri (TT)
stars, young (1-2 Myr) brown dwarfs possess protoplanetary disks of dust
and gas \citep[e.g.][]{Luhman2007} that are the environment of planet formation
\citep[e.g.][]{Apai2005}.  Indeed, the recent discovery of the TRAPPIST-1 exoplanet system
at the sub-stellar boundary \citep{Gillon2016, Gillon2017}

suggest that brown dwarfs may be capable of forming planetary systems.  Observing
protoplanetary disks in this regime allows us to observe the lower-boundary
conditions of planet formation.

The conditions in protoplanetary disks are strongly dependent on stellar mass.
In particular, scaling laws have been identified between host star mass and
protoplanetary dust disk mass \citep[e.g.][]{Pascucci2016}.
Planet formation models typically adopt these scaling laws to predict the
planets that may form around low-mass stars 
\citep[e.g.][]{Raymond2007,Mordasini2012}.
However, disks around the lowest-mass stars may not be scaled down versions of their
higher-mass counterparts. Disks around brown dwarfs show differences in their
dust and gas evolution \citep{Pascucci2009}. There are also indications that
disks around brown dwarfs have a different disk geometry \citep[e.g.][]{SzHucs2010,Daemgen2016}, though this is not always found \citep{Harvey2010,Mulders2012}.

\floattable
\begin{deluxetable}{clcllllccc}
\onecolumngrid

    \tablecaption{Source Properties\label{table-properties}}

\tablehead{
\colhead{2MASS} &
    \colhead{Common} &
    \colhead{Region} &
    \colhead{SpTy} &
    \colhead{Ref.} &
    \colhead{$A_\text{v}$} &
    \colhead{log $L_{\text{bol}}$} &
    \colhead{Ref.} &
    \colhead{$T_{\text{eff}}$} &
        \colhead{$M_{*}$} \\
\colhead{} &
    \colhead{Name} &
    \colhead{} &
    \colhead{} &
    \colhead{} &
    \colhead{} &
    \colhead{($L_\sun$)}
    & \colhead{}
    & \colhead{(K)}
    & \colhead{(M$_\sun$)}
}
\startdata
    J04141760+2806096  & \OID{1342268646} & Taurus & M5  & L17 & 4.6                   & -0.7  & R10 & 3125 & 0.19\\
    J04193545+2827218  & \OID{1342263516} & Taurus & M5.25 & L17 & 0.0                   & -0.9  & R10 & 3090 & 0.14\\
    J04233539+2503026  & \OID{1342264241} & Taurus & M7 & L17 & 2.0                   & -0.72 & L09 & 2880 & 0.16\\
    J04295950+2433078  & \OID{1342265469} & Taurus & M5  & L17 & 4.6                   & -0.7  & R10 & 3125 & 0.19\\
    J04381486+2611399  & \OID{1342265470} & Taurus & M7.25 & L17 & 3.5                   & -2.3  & R10 & 2846 & 0.05\\
    J04382134+2609137  & \OID{1342264239} & Taurus & M5  & L17 & 2.0                   & -1.15\tablenotemark{a} & H08 & 3125 & 0.14\\
    J04393364+2359212  & \OID{1342263934} & Taurus & M5    & L17 & 1.3                   & -1.0  & R10 & 3125 & 0.18\\
    J04394748+2601407  & \OID{1342264240} & Taurus & M7    & L17 & 4.4                   & -1.0  & R10 & 2880 & 0.11\\
    J11062554--7633418 & \OID{1342263489} & Cha~I  & M5.25 & L07 & 3.58\tablenotemark{b} & -1.28 & L07 & 3088 & 0.14\\ 
    J11071668--7735532 & \OID{1342263459} & Cha~I  & M7.75 & L07 & 0.0\tablenotemark{b}  & -1.82 & L07 & 2765 & 0.04\\
    J11105597--7645325 & \OID{1342263492} & Cha~I  & M5.75 & L07 & 2.91\tablenotemark{b} & -0.89 & L07 & 3021 & 0.14\\ 
\enddata
\tablecomments{A distance of 137 and 162\,pc is assumed for the Taurus (R10) and Chamaeleon~I sources (L07).}
\tablenotetext{a}{GM Tau luminosity from D14}

\tablenotetext{b}{\,L07 report the extinction at J band. From those values we computed the $A_{\rm V}$ in this table using the interstellar extinction law in Mathis~(1990) and an Rv of 3.1.}
\tablerefs{
    \cite{Davies2014} (D14);
    \cite{Herczeg2008} (H08);
    \cite{Luhman2007} (L07); \cite{Luhman2009} (L09); \cite{Luhman2017} (L17);
    \cite{Rebull2010} (R10)
}
\end{deluxetable}

Because brown dwarfs are significantly fainter than TT stars, only limited observations
are present to directly constrain their dust masses \citep[e.g.][ and
references therein]{Mohanty2013,Pascucci2016}.  Several studies have used
Herschel far-infrared photometry
\citep[e.g.][]{Harvey2012,vanderPlas2016,Daemgen2016} and ALMA
\citep[e.g.][]{vanderPlas2016} to estimate dust masses. One parameter that has
remained largely unexplored in these studies is the dust disk outer radius. From a
theoretical perspective, disks around brown dwarfs are predicted to form with smaller
gas disks \citep{Bate2012} and experience more efficient radial drift leading
to smaller dust disks \citep{Pinilla2013}.  Observations with ALMA of a handful
of spatially resolved brown dwarf dust and gas disks range in size from
20~au to 140~au \citep[][]{Ricci2013,Ricci2014,Testi2016}, smaller than 
TT disks (22--440~au with a mean size of 165~au) \citep{Isella2009,Andrews2010,Guilloteau2011}.

In this work, we explore the impact of smaller disk radii around brown dwarfs
and very-low mass stars (hereafter VLMOs) on
observables related to the dust and the gas in those disks.  We present a
Herschel survey of \OI{} for 11 VLMOs probing the disk
gas in \S~\ref{sect:survey} and \S~\ref{sect:oidet}.  Detailed thermo-chemical
models of protoplanetary disks have shown that this transition can be used as
an order of magnitude disk mass estimator \citep[e.g.][]{Woitke2010}, hence a
large survey toward T Tauri stars \citep{Dent2013} was carried out.  We perform radiative
transfer modeling of the spectral energy distribution (SED) of these sources in
\S~\ref{sect:individual} starting from the zero-order assumption that the disk
geometry and dust opacity are stellar-mass independent. We also run two
radiative transfer grids with and without a stellar-mass dependent disk radius
to quantify the difference in disk temperature in
\S~\ref{sect:methods_grid_models}.  We examine the underluminosity of \OI{} in
our VLMO disks in \S~\ref{sect:faint}, while in \S~\ref{sect:small} we compare
the likelihoods of small and large disk models by testing model grids against
the SEDs of our VLMO sample. We also show how smaller disk radii lead to higher
disk temperatures than previously assumed and impacts estimates of the dust
disk mass. 

\section{Observations and Data Reduction}\label{sect:survey}

\subsection{Sample}\label{sect:sou}

Our sample comprises 11 very low-mass stars and brown dwarfs from two nearby
star-forming regions, Taurus and \ChaI{} ($d=162$\,pc, \citealt{Luhman2008}).
Because we use stellar parameters from \cite{Rebull2010} for the majority of
our Taurus sources, we adopt their distance of 137~pc for the Taurus
star-forming region \citep{Torres2007,Torres2009}.  Ten of the sources were
selected by us and observed with the Herschel/PACS spectrograph as part of
Proposal ID \texttt{OT2\_ipascucc\_2}, see \S~\ref{sect:obs} for more details.
CIDA-1 is the only other very low-mass star observed with the same setting
(Proposal ID \texttt{OT1\_ascholz\_1}), hence we include it in our analysis.

We chose to focus on the Taurus and \ChaI{} star-forming regions because the
low-mass end of the stellar population is well characterized and, unlike
$\rho$~Oph, suffer from minimal contamination from extended cloud emission.  In
addition to being very low-mass stars/brown dwarfs, our targets were selected to posses
a dust disk (based on infrared excess emission) and to have a flux density at
24\,\micron{} ($F_{24}$) greater than 50\,mJy
\citep{Guieu2007,Luhman2008a,Rebull2010}. CIDA-1 also fits these criteria.  The
24\,\micron{} flux limit was applied to ensure bright 70\,\micron{} fluxes
($F_{70}$), greater than $\sim 70$\,mJy, based on the empirical relation
$\nu_{70} F_{70} \sim 0.5 \times \nu_{24} F_{24}$ from the VLMO disk sample of
\citet{Harvey2012}.  Because of the known positive correlation between the
far-infrared continuum and the \OI{} emission line
\citep[e.g.][]{Howard2013,Keane2014}, our target selection criteria were
intended to maximize the detection of the [OI] forbidden line (see also
\S~\ref{sect:obs}).

Table \ref{table-properties} provides the main properties of our targets.
Spectral types are converted to effective temperatures for each object using the relationship given in \cite{Luhman2003}.
Because this relationship deviates from linearity between spectral
types M6 and M9, quadratic interpolation was used to estimate temperatures for
non-integer spectral types.  Stellar masses were computed by comparing each
source position in the Hertzsprung-Russell diagram to the \citet{Baraffe1998} evolutionary tracks.  While this is a standard approach
and the only one available for our sample, one should keep in mind that
model-inferred masses have large uncertainties of 50-100\% below $1\,M_\sun$ as
demonstrated by \citet{Stassun2014} using a sample of eclipsing binary systems. 

Two of our sources, FU~Tau and Hn~13, are multiple.  FU Tau has an M9.25
companion at a separation of $5\farcs7$, and may have an additional
spectroscopic binary companion \citep{Luhman2009}.  Hn~13 has a $0\farcs13$
near-equal mass companion (imaged by \cite{Ahmic2007}).

\subsection{Herschel/PACS Spectroscopy}\label{sect:obs}

Our sample was observed in February and March 2013 with the ESA
\textit{Herschel Space Observatory} \citep{Pilbratt2010} Spectrograph \citep[PACS;][]{Poglitsch2010} which provides an IFU with a $5\times5$ array of
spaxels, each with a pixel size of 
$9\farcs4$. The sources were observed
under two different programs (\texttt{OT2\_ipascucc\_2} 10 sources and
\texttt{OT1\_ascholz\_1} for CIDA-1).  Both programs used the standard
point-source spectroscopy line scan mode with chop/nod, the chopper throw set
to small and the standard faint line mode selected (high grating sampling
density).  Spectra were centered on the \OI{} line and covered
from 62.93 to 63.43 $\mu$m.  The red channel observations were a bonus and
covered from 188.77 to 190.35 $\mu$m.

We set the exposure times for the ten \texttt{OT2\_ ipascucc\_2} sources by
extrapolating the 63\,\micron{} continuum-line relationship from
\citet{Howard2013} down to a minimum flux density of 70\,mJy (see
\S~\ref{sect:sou}) and found a corresponding \OI{} line flux of
4\e{-18} W/m$^2$.  The exposure time was set to 16,420 seconds to achieve a
3$\sigma$ detection on this line, i.e. a sensitivity of 1.4\e{-18} W/m$^2$,
about three times better than that achieved for TT stars with the GASPS survey
\citep{Dent2013}.  CIDA-1, the brightest source in our sample and observed
under \texttt{OT1\_ascholz\_1}, has a shorter exposure time of 9,868 seconds.
The particulars of the observations including the \textit{Herschel} observation
identification numbers (ObsIDs)  are listed in Table \ref{table-observations}.

\begin{deluxetable}{l ccc}

    \tablecaption{Herschel/PACS Sample and Observations \label{table-observations}}

\tablehead{
    \colhead{Target Name} & \colhead{ObsID} & \colhead{Obs Date} & \colhead{Duration ($s$)}
}
\startdata
    CIDA-1            & 1342268646 & 2013-03-25 & 9868.0    \\
    FR~Tau            & 1342263516 & 2013-02-12 & 16420.0   \\
    FU~Tau~A          & 1342264241 & 2013-02-25 & 16420.0   \\
    CFHT-20           & 1342265469 & 2013-02-16 & 16420.0   \\
    J04381486         & 1342265470 & 2013-02-16 & 16420.0   \\
    GM~Tau            & 1342264239 & 2013-02-25 & 16420.0   \\
    J04393364         & 1342263934 & 2013-02-19 & 16420.0   \\
    CFHT-4            & 1342264240 & 2013-02-25 & 16420.0   \\
    ESO~H$\alpha$~559 & 1342263489 & 2013-02-11 & 16420.0   \\
    Cha~H$\alpha$~1   & 1342263459 & 2013-02-10 & 16420.0   \\
    Hn~13             & 1342263492 & 2013-02-11 & 16420.0   \\
\enddata

\end{deluxetable}

The data were reduced from level 0 using the \textit{Herschel}
Interactive Processing Environment \citep[HIPE;][]{Ott2010} version 14.0.1 with
calibration set product 72.  Because the targets are all faint point-sources,
the ``ChopNodBackgroundNormalization" (version 1.38.4.3) pipeline script was
used for reduction.  This pipeline flags bad data (e.g. mechanism movements,
saturation, overly-noisy or bad pixels), applies signal corrections (e.g.
non-linearities, crosstalk, transients, capacitance ratios), performs
wavelength and flux calibration, and corrects data for spacecraft velocity.
For the final cube rebinning, the wavelength grid was sampled using
\texttt{oversample=2} and \texttt{upsample=4}. Proper flatfielding was verified
interactively from within HIPE for each target.

\begin{figure*}[tbh]
    \makebox[\textwidth]{%
        \includegraphics[width=0.33\textwidth]{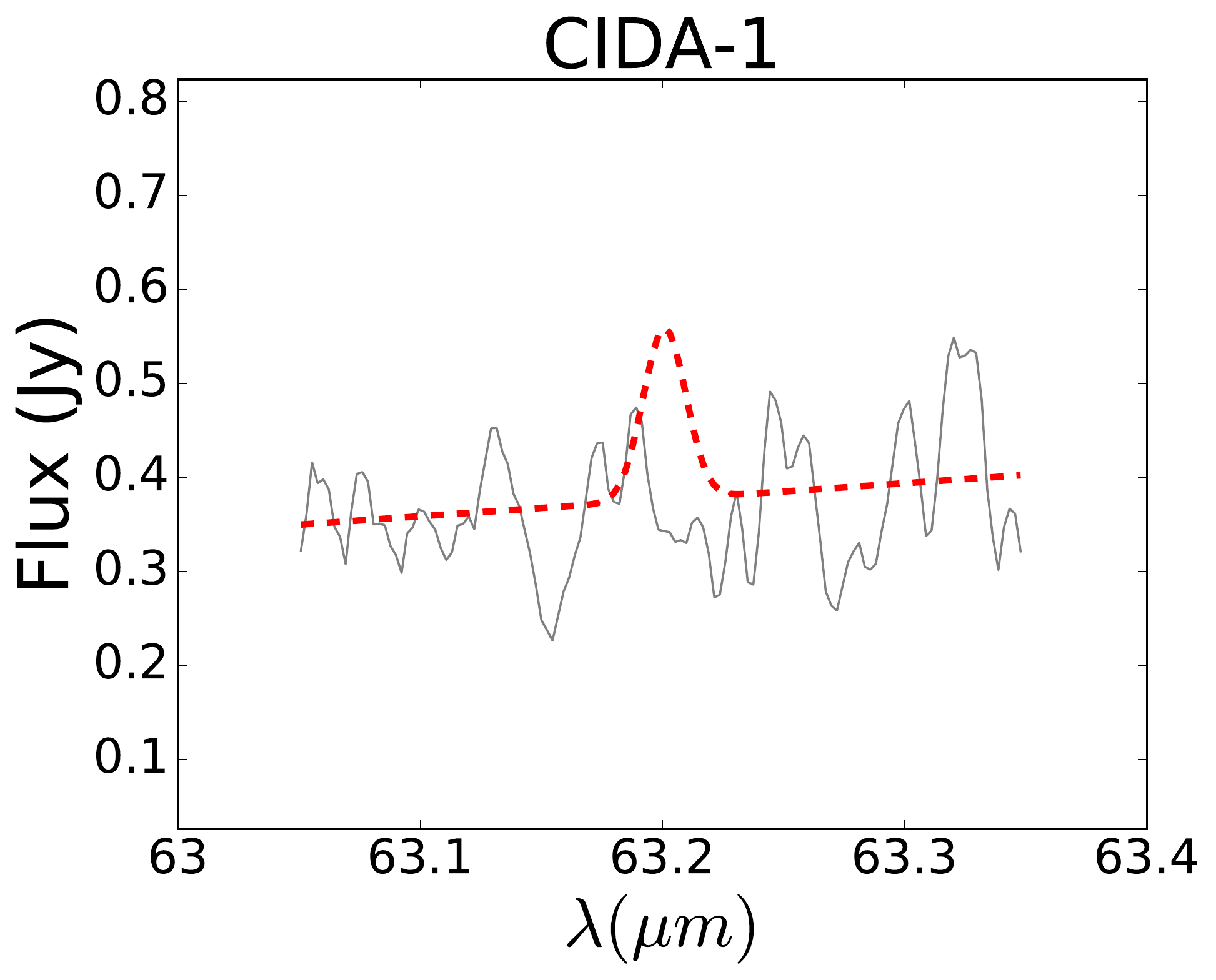}%
        \hfill    
        \includegraphics[width=0.33\textwidth]{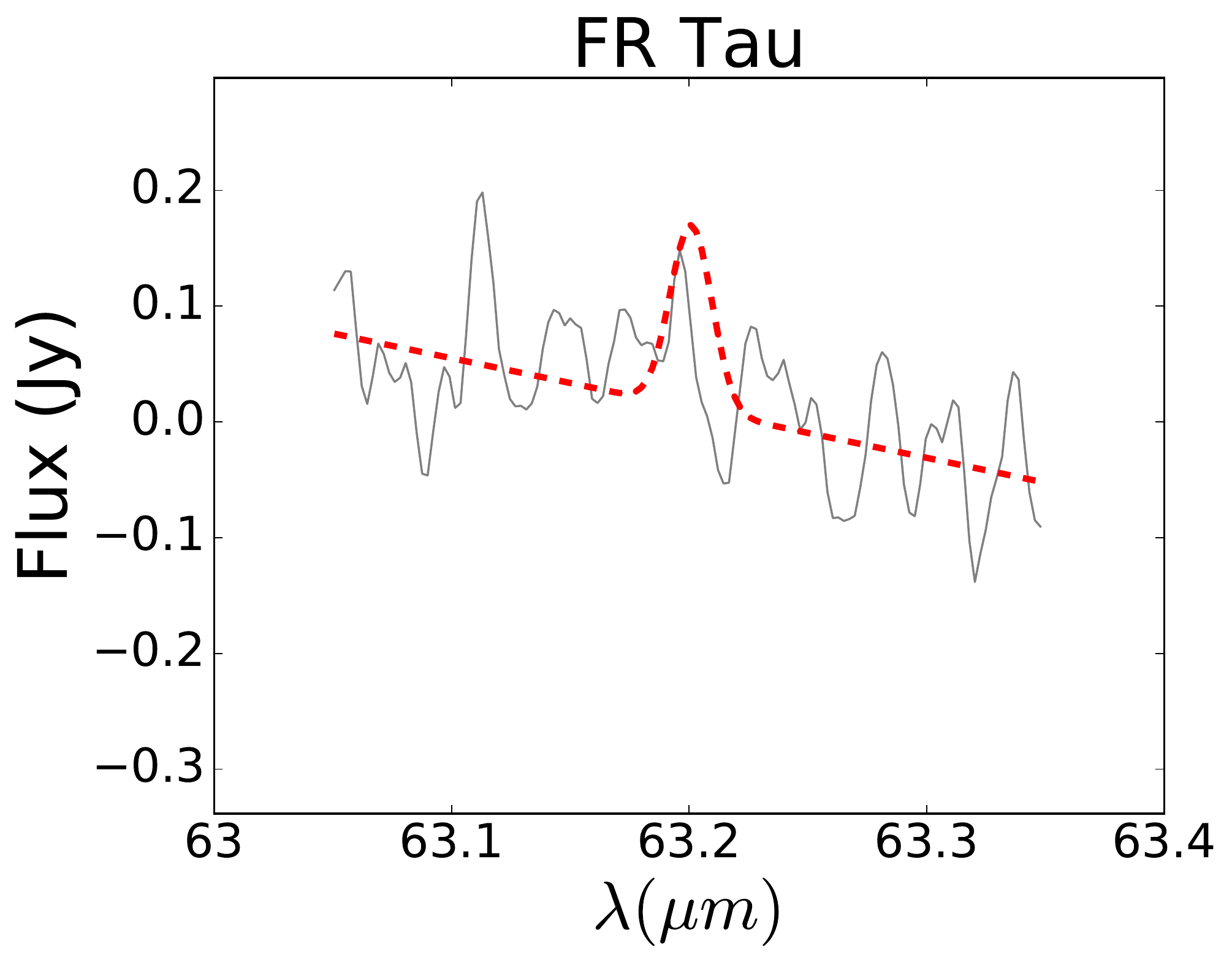}%
        \hfill    
        \includegraphics[width=0.33\textwidth]{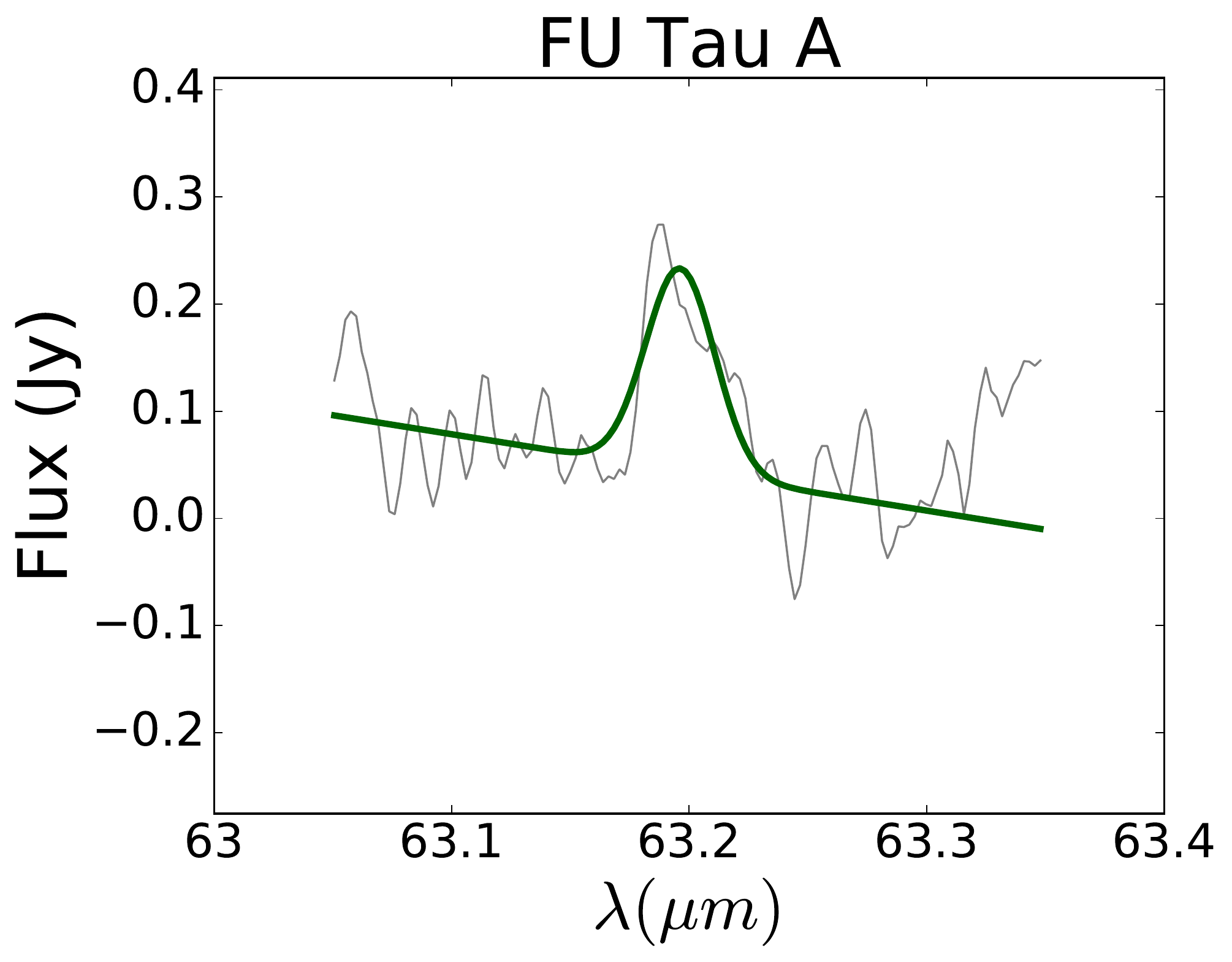}%
    }\\
        \makebox[\textwidth]{%
        \includegraphics[width=0.33\textwidth]{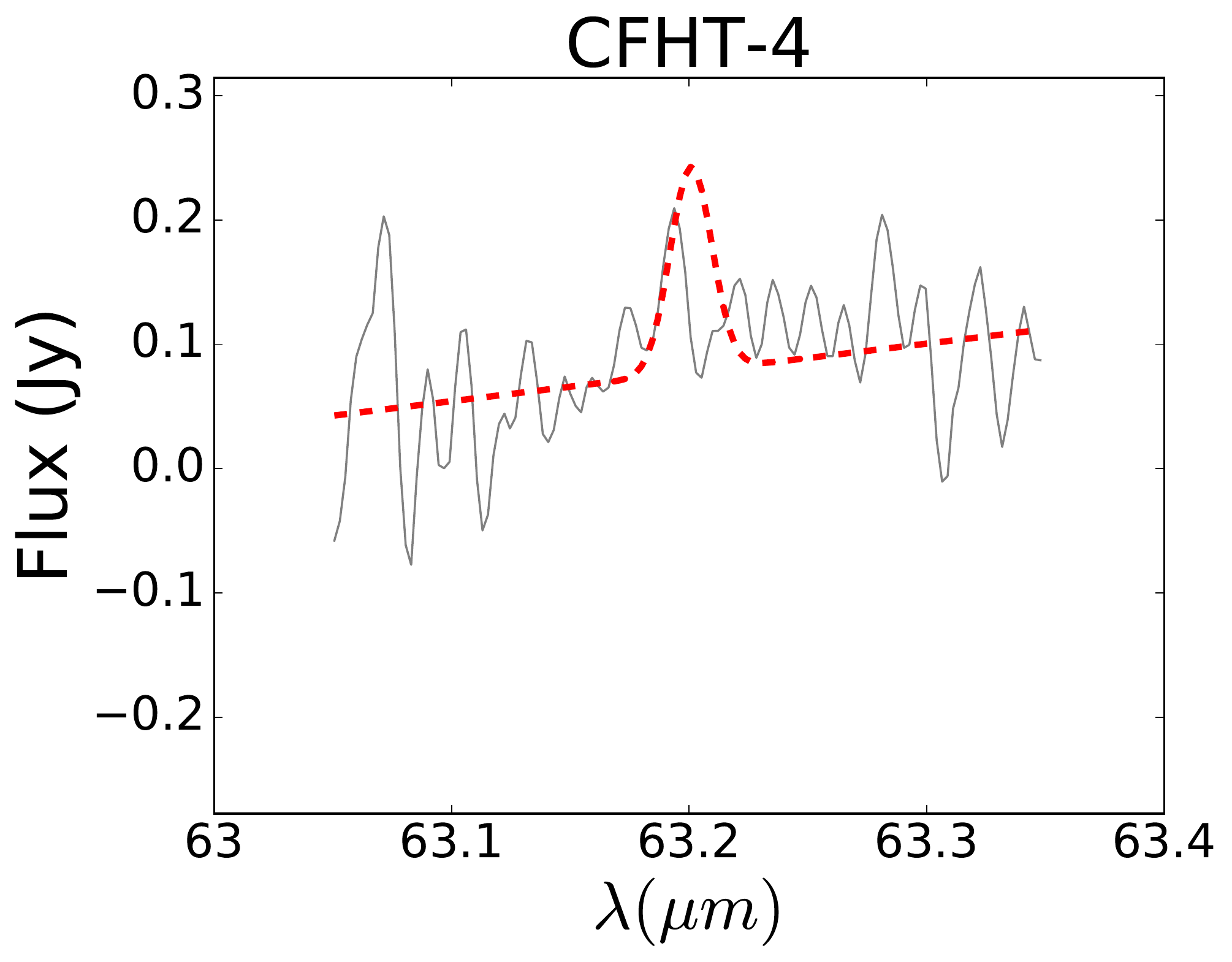}%
        \hfill    
        \includegraphics[width=0.33\textwidth]{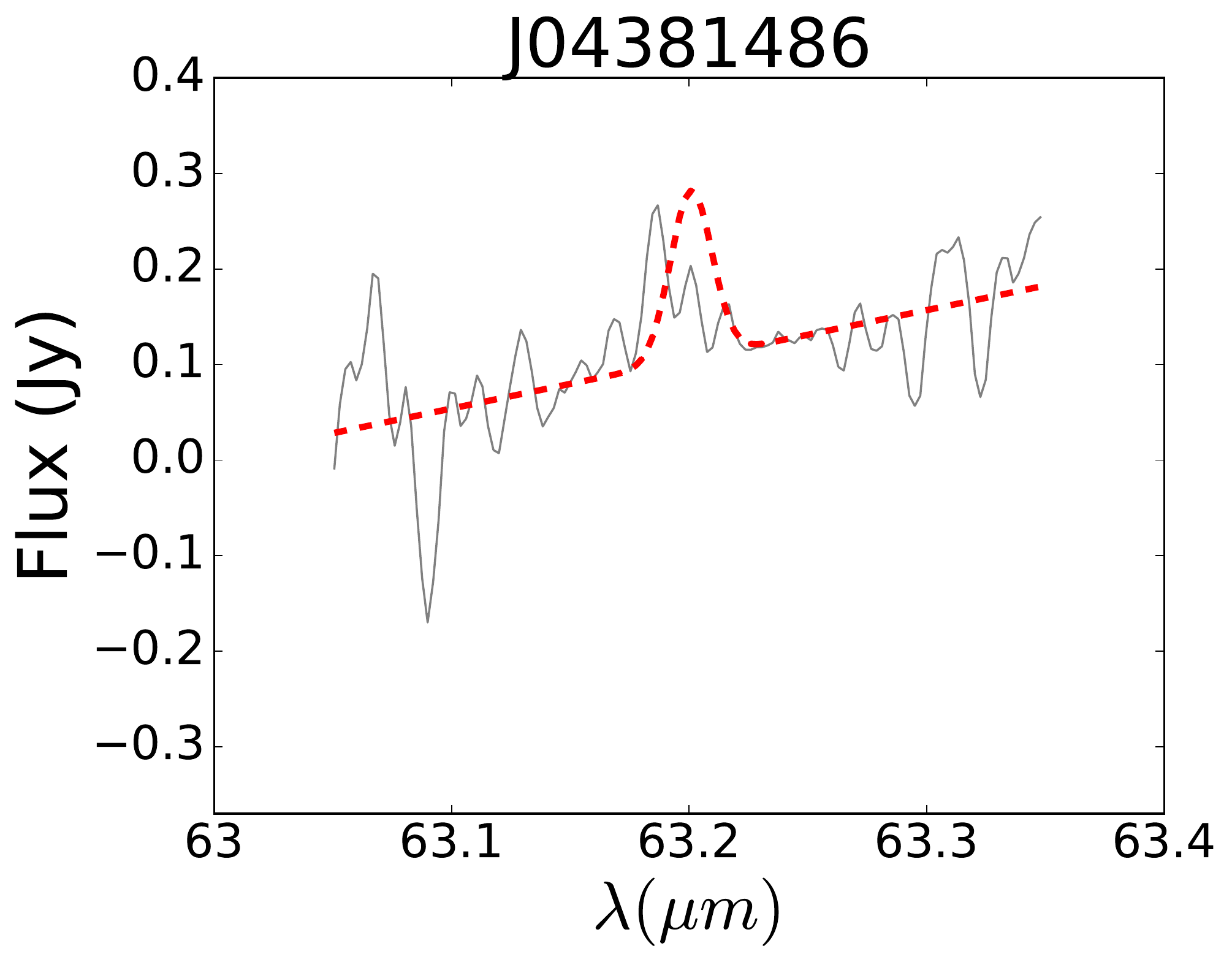}%
        \hfill    
        \includegraphics[width=0.33\textwidth]{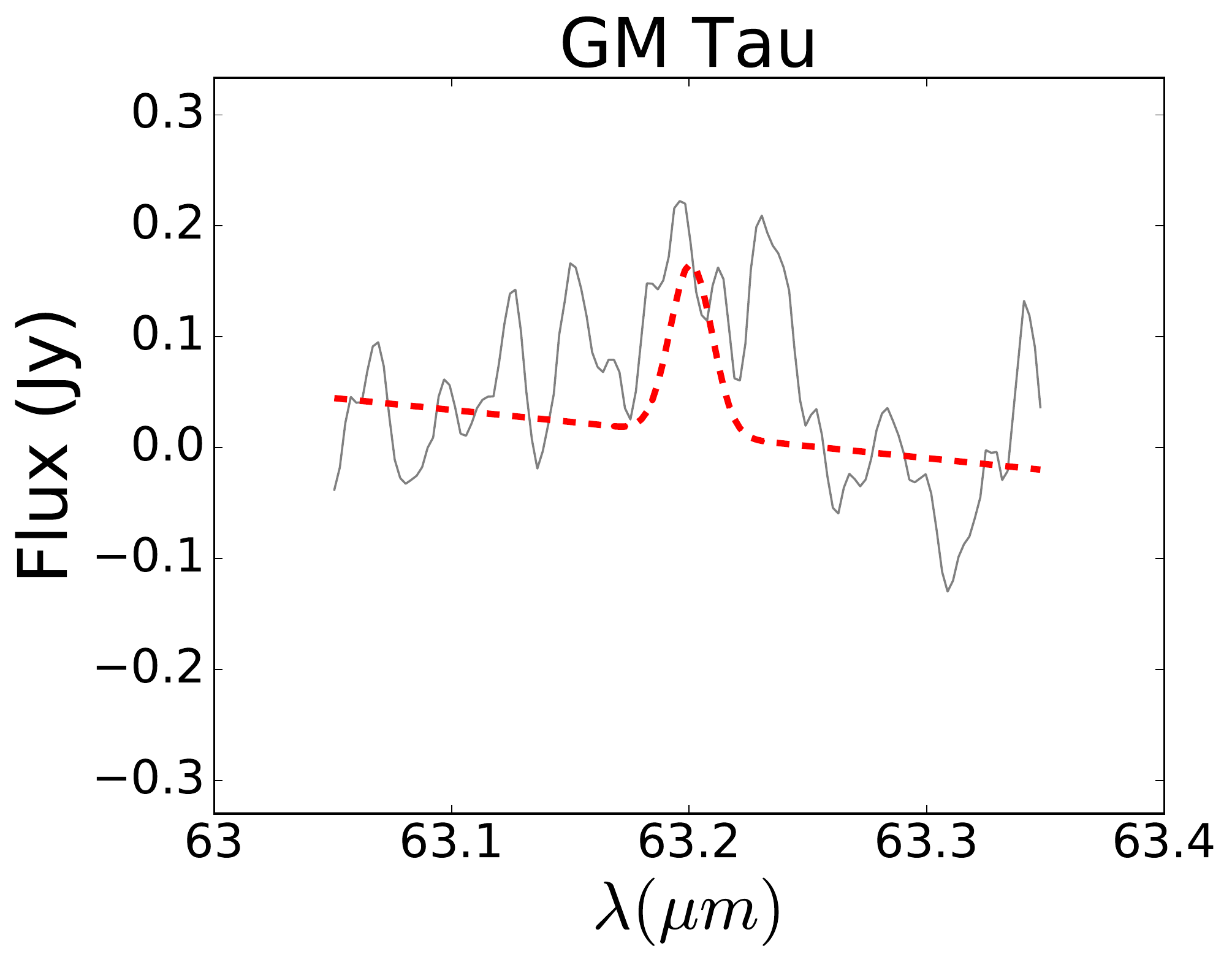}%
    }\\
        \makebox[\textwidth]{%
        \includegraphics[width=0.33\textwidth]{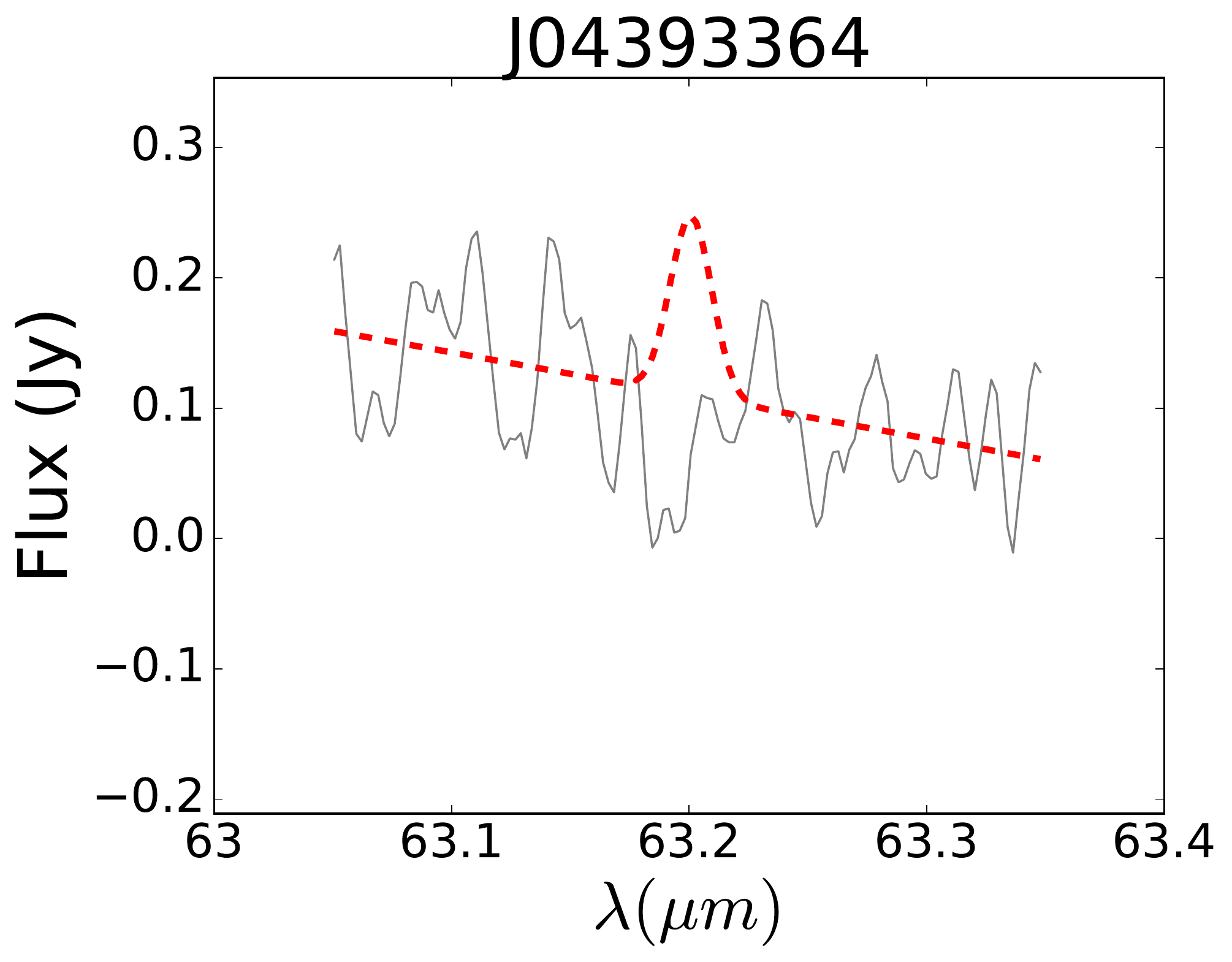}%
        \hfill    
        \includegraphics[width=0.33\textwidth]{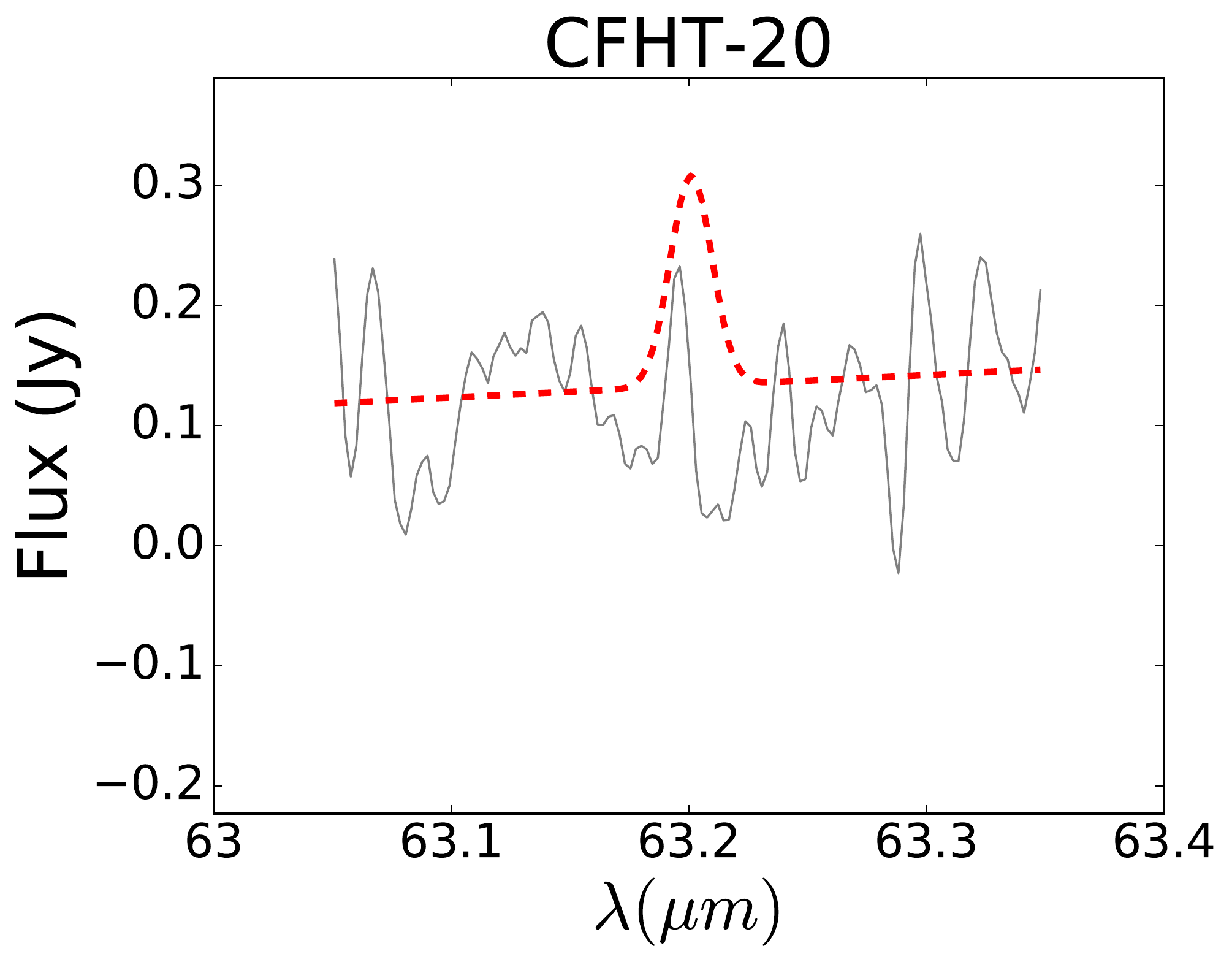}%
        \hfill    
        \includegraphics[width=0.33\textwidth]{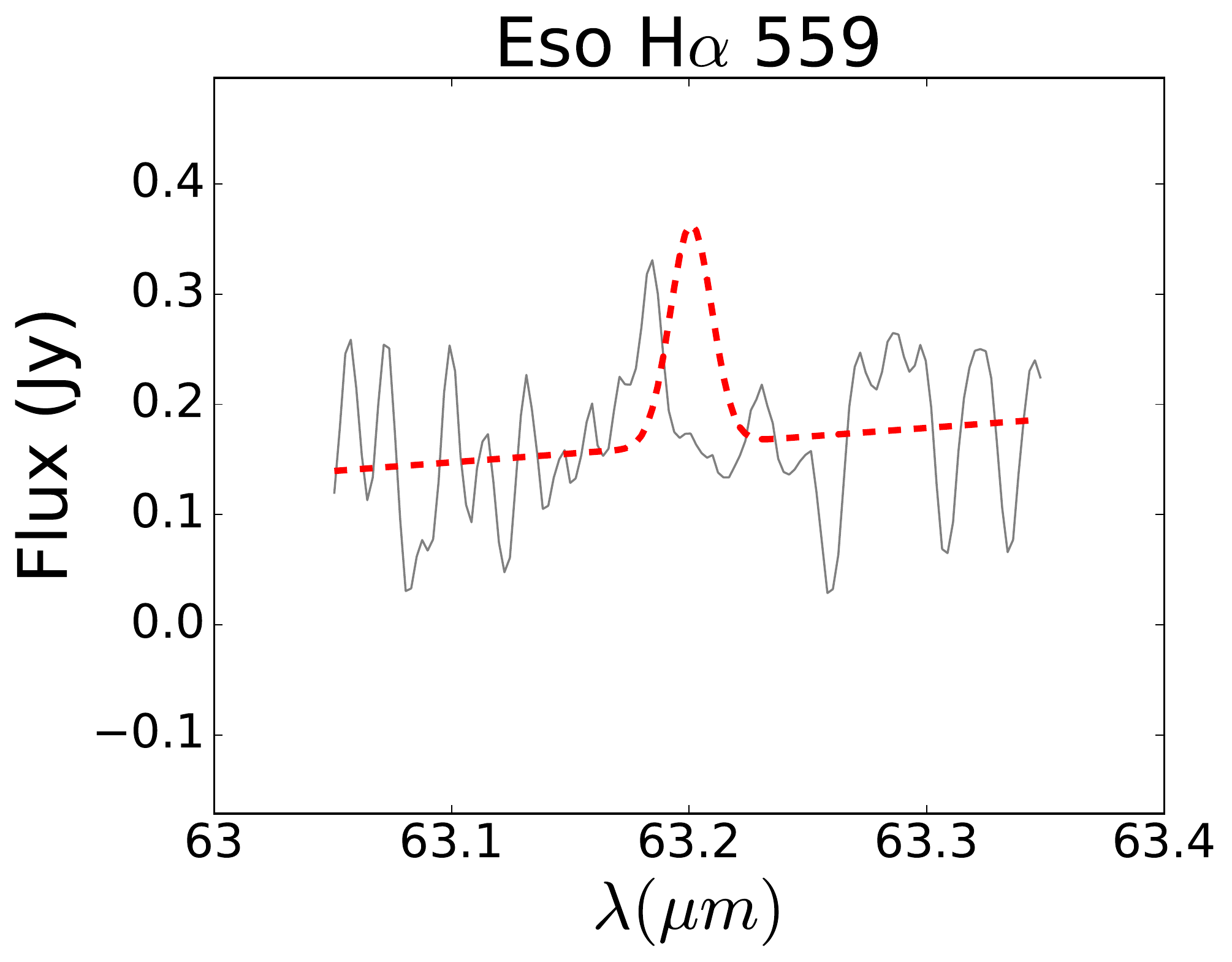}%
    }\\%
        \makebox[\textwidth]{%
        \includegraphics[width=0.33\textwidth]{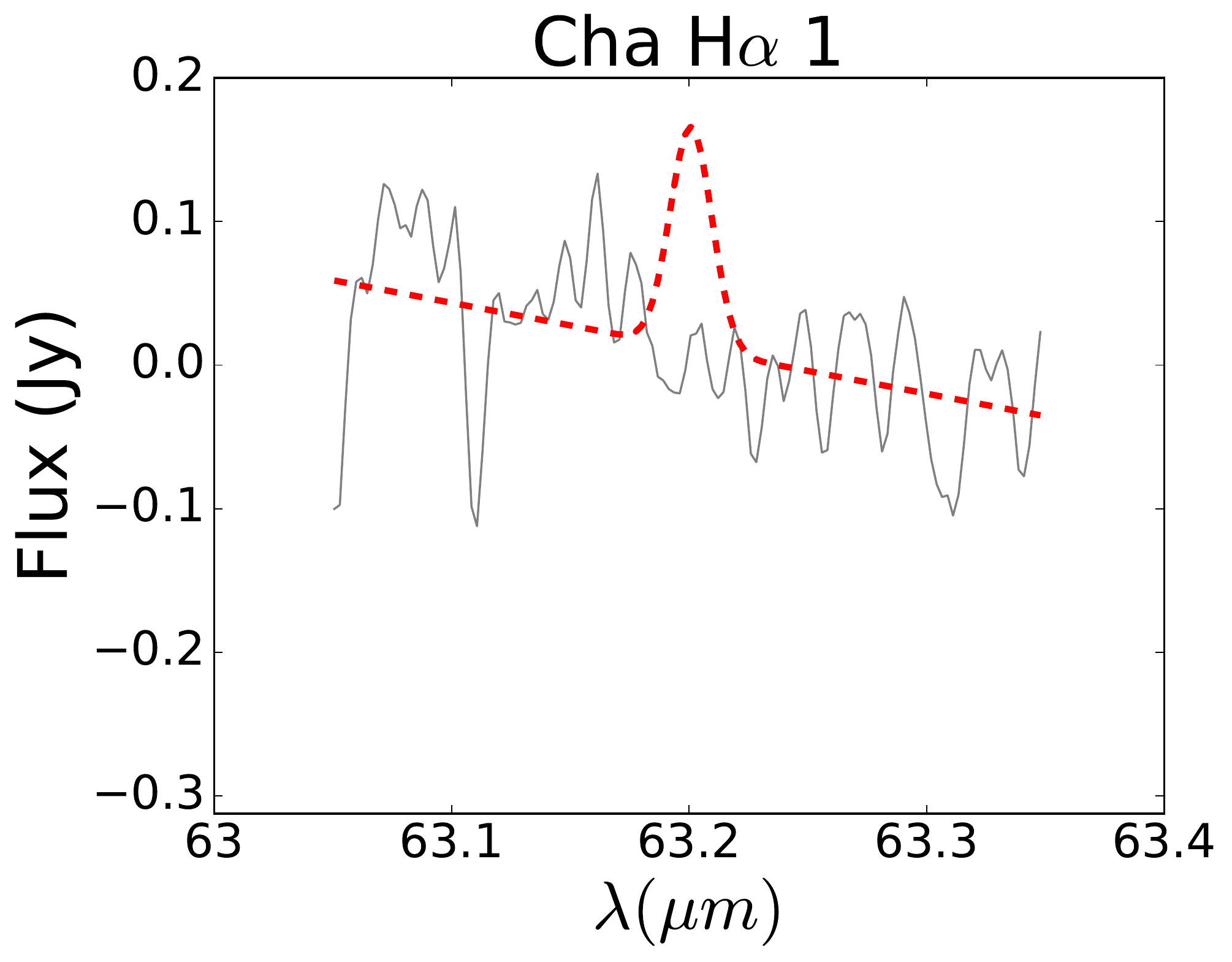}%
        \includegraphics[width=0.33\textwidth]{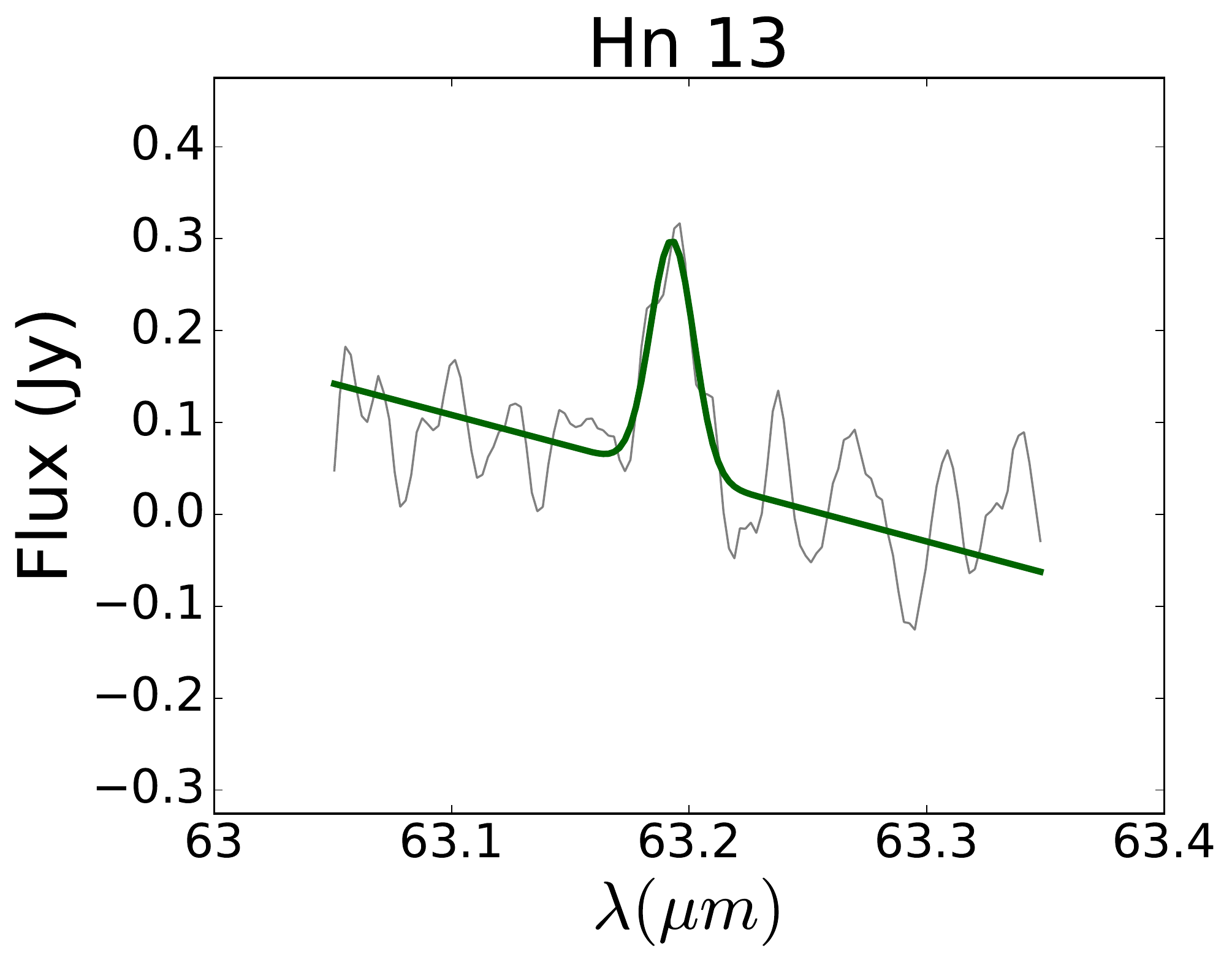}%
    }%
    \caption{
        Herschel/PACS spectroscopy measurements of the wavelength region around the \OI{} emission line (solid gray line).  
        Greater than $3\sigma$ Gaussian fits (detections) are overplotted as solid green lines.
        Hypothetical $3\sigma$ upper limits (FWHM of unresolved line) are shown as red dashed lines.
    }
    \label{figure-blue_spectra}
\end{figure*}

Because telescope jitter and pointing errors can result in flux beyond the
central spaxel, the appropriate spectra output from the task
\texttt{extractCentralSpectrum} must be chosen as described in the \textit{PACS
Data Reduction Guide: Spectroscopy} (section 8.4.1).  The central spaxel with
the application of the c1-to-total point-source correction was used to extract
the final spectrum.  To further test this choice the spectra of neighboring
spaxels were examined for extended emission using the ``comboplot" output.  All
neighboring spaxels were confirmed to have no signal above the rms.  For all of
our sources the central spaxel (\texttt{c1}) spectra had the best SNR and a
signal that was comparable to those extracted from the 3x3 corrected spectra.
The absolute flux calibration uncertainty for the PACS spectrograph at
$\sim$63\,\micron{} is estimated to be 10\% (PACS Observer Manual;
HERSCHEL-HSC-DOC-0832, Version 2.5.1).  The final spectra for all of our
targets are shown in Figure~\ref{figure-blue_spectra}.

\section{Analysis}

Our analysis is separated into 3 sub-sections.  In \S~\ref{sect:oidet} we
present our findings for the 63\,\micron{} [OI] line emission and continuum,
and 189\,\micron{} continuum for 11 \HSO observations.  The \OI{} emission
informs us about the gas
content, especially in the disk atmosphere \citep[e.g.][]{Woitke2010}, while the continuum emission about the dust in the disk.

This is followed by
two separate sets of radiative transfer modeling schemes.  The first,
\S~\ref{sect:individual}, investigates VLMO dust disk properties by fitting our
source SEDs.  \S~\ref{sect:methods_grid_models} describes the grid of models we
used to analyze the relationship between the disk outer radius and the derived
disk mass.

\floattable
\begin{deluxetable}{l ccccc}
\onecolumngrid

    \tablecaption{Observed continuum flux densities and line fluxes.\label{table-fluxes}}

\tablehead{
    \colhead{Source} &
    \colhead{$F_{\rm [OI]}$}&
    \colhead{$F_{63}$} &
    \colhead{$F_{189}$} &
    \colhead{$F_{70}$} &
    \colhead{Ref.} \\
    \cmidrule(lr){2-4}
    \cmidrule(lr){5-6}
    \colhead{} &
    \colhead{($10^{-18}$ W/m$^2$)}&
    \colhead{(mJy)} &
    \colhead{(mJy)} &
    \colhead{(mJy)} &
    \colhead{($F_{70}$)} 
}
\startdata
    CIDA-1            & $\le 3.01$       & $358.7  \pm 50.9$ & $270.4 \pm 83.3$  & $266 \pm 2$ & B14 \\
    FR~Tau            & $\le 2.62$       & $\le 120.7$       & $\le 145.4$   & $46 \pm 3$  & B14 \\
    FU~Tau~A          & $5.17  \pm 1.04$ & $\le 98.9$        & $\le 131.3$  & $86 \pm 17$ & R10 \\
    CFHT-20           & $\le 2.89$       & $\le 127.6$       & $\le 154.5$   & $128 \pm 4$ & B14 \\
    J04381486         & $\le 2.91$       & $\le 132.0$       & $\le 152.5$   & $95 \pm 2$  & B14 \\
    GM~Tau            & $\le 2.54$       & $\le 121.5$       & $\le 66.8$  & $36 \pm 2$  & B14 \\
    J04393364         & $\le 2.28$       & $100.3  \pm 32.4$ & $\le 140.0$   & $70 \pm 1$  & B14\\
    CFHT-4            & $\le 2.73$       & $\le 110.6$       & $\le 161.9$ & $109 \pm 5$ & B14\\
    ESO~H$\alpha$~559 & $\le 3.30$       & $174.0  \pm 45.6$ & $317.7 \pm 42.2$  &           & \\
    Cha~H$\alpha$~1   & $\le 2.55$       & $\le 112.5$       & $172.6 \pm 40.0$  &           & \\
    Hn~13             & $4.35  \pm 1.30$ & $\le 114.0$       & $\le 76.7$  &           & \\
\enddata
\tablecomments{$3\sigma$ upper limits are reported for non-detections.}                                                                 

\end{deluxetable}

\subsection{[OI] Line and Continuum Detections}\label{sect:oidet}

To identify \OI{} detections, we smooth the spectrum using a
uniform filter (width of 3 resolution elements) before fitting each spectrum within $\pm
0.1$\,\micron{} of the line using a Levenberg-Marquardt algorithm assuming a
Gaussian for the line profile and a first-order polynomial for the continuum.
The 1$\sigma$ uncertainties on the line fluxes are evaluated from the standard
deviation of the pixels in the spectrum minus the best-fit model.  We consider
a line to be detected when its flux is greater than 3 times the 1$\sigma$
uncertainty. In case of non-detections, we fit the same spectral range with a
first-order polynomial and we provide in Table~\ref{table-fluxes} the 3$\sigma$
upper limits from the rms in the baseline-subtracted spectrum using a Gaussian
with a width equal to that of an unresolved line (FWHM of 98\,km/s at
63.18\,\micron ). Sources are considered detected in the continuum if the S/N
at 63.2\,\micron{} is greater than 3. 

Following this approach we detect the \OI{} line of two sources in our sample
(toward FU~Tau~A and Hn~13) and the continuum of three bright disks (CIDA-1,
J04393364, and ESO~H$\alpha$~559). Two of these bright disks (CIDA-1 and
ESO~H$\alpha$~559) are also detected in the continuum at 189\,\micron{}. FU~Tau~A has a known
molecular outflow \citep{Monin2013}, and it is possible that the outflow
contributes to the \OI{} line, in which case the flux we report in
Table~\ref{table-fluxes} would be an upper limit for the disk emission.

\citet{Riviere-Marichalar2016} analyzed 362 Herschel sources, which included
our 11 targets, using HIPE 12.0.  Our results are consistent with
\citeauthor{Riviere-Marichalar2016} except for FU~Tau~A where we report an
$\sim 5 \sigma$ \OI{} detection ($5.17  \pm 1.04\times10^{-18}~W/m^2$) while
\citeauthor{Riviere-Marichalar2016} report a $3 \sigma$ upper limit of
($<3\times10^{-18}~W/m^2$).

All of the eight Taurus sources were also observed with the PACS photometer by
\cite{Bulger2014}.  Literature 70\,\micron{} flux densities are reported in
the last column of Table~\ref{table-fluxes}.  Two of the sources (FR~Tau and
GM~Tau) are fainter than the 70\,mJy flux density estimated from the
24\,\micron{} photometry available at the time of the proposal submission.
Except for CIDA-1, our continuum values and upper limits are consistent with
the literature values within 1$\sigma$ of the uncertainties we quote. In the
case of CIDA-1 the 63 and 70\,\micron{} flux densities are within 2$\sigma$. It
is possible that source variability further contributes to the flux discrepancy
for this source as 70\,\micron{} variability can be close to $\sim 20$\% even
for non-embedded young low-mass stars (Class~II SEDs, see \citealt{Billot2012}).
Given the better sensitivity of the PACS photometer and smaller uncertainty of
the 70\,\micron{} photometric values, we will use them, when available, in the
analysis that follows.

\begin{figure*}
    \includegraphics[width=\textwidth]{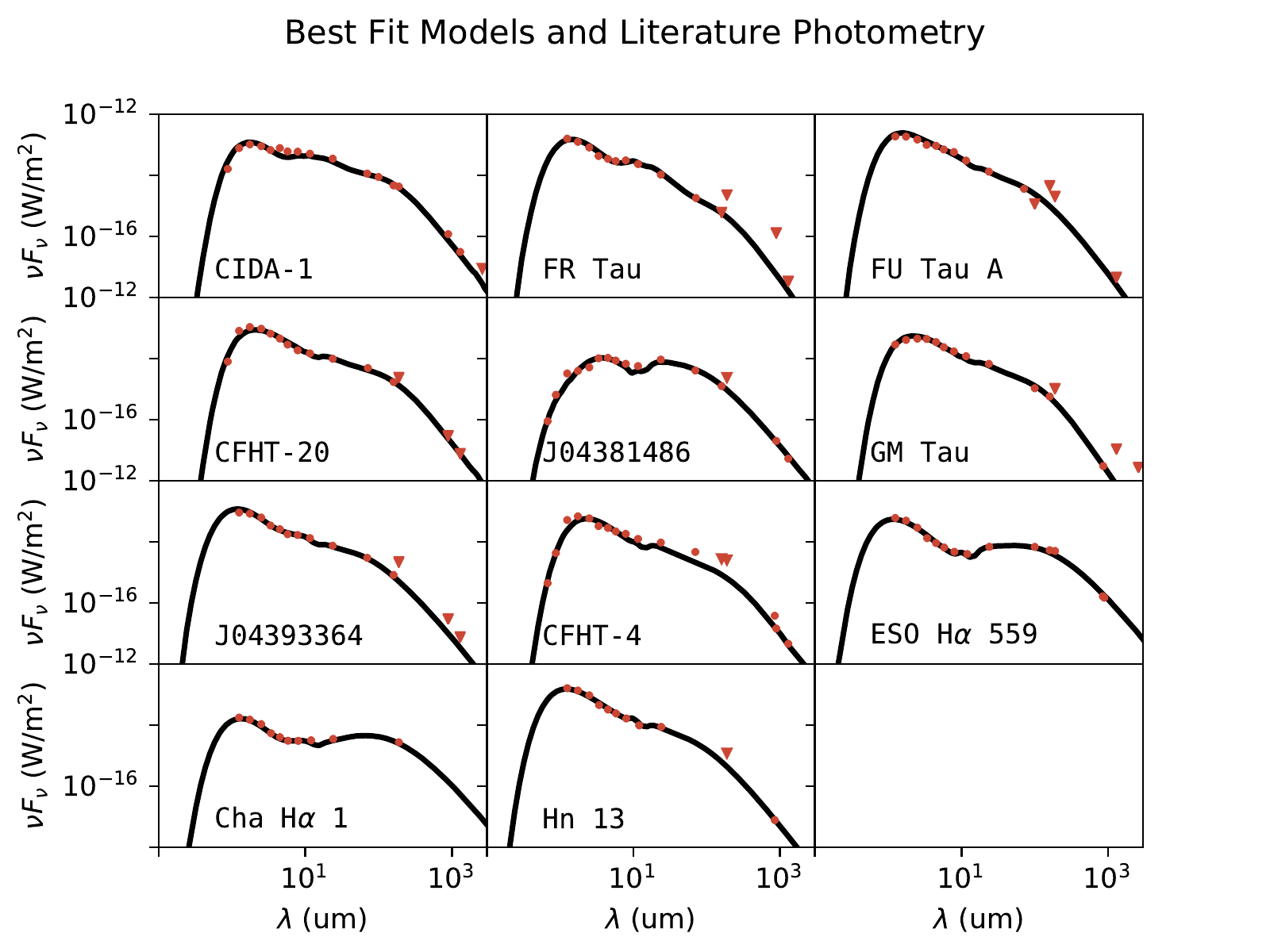}
    \caption{
        \label{figure-seds}
        SEDs of our Herschel sources.  Orange: photometric fluxes from literature; triangles: upper limits.  The best-fit radiative transfer models for each source are displayed for each target (black line).
    }
\end{figure*}

\subsection{Continuum Radiative Transfer Modeling of Individual Sources}\label{sect:individual}

\floattable

\begin{deluxetable}{|l|p{8cm}|p{8cm}|p{5cm}|}
\onecolumngrid
\tablecaption{Source photometry.\label{table-source_photometry}}
\tablehead{
 \colhead{Source} &
 \colhead{Wavelength} &
 \colhead{Flux} &
 \colhead{Reference} \\
 \colhead{} &
 \colhead{($\micron$)} &
 \colhead{(mJy)} &
 \colhead{}
}
\rotate
\startdata
\OID{1342268646} & 0.88, 1.25, 1.75, 2.50, 3.35, 4.49, 5.73, 7.87, 11.56, 23.67, 70.00, 100.00, 160.00, 189.00, 890.00, 1300.00, 2600.00 & 0.0047, 0.033, 0.06, 0.075, 0.075, 0.12, 0.12, 0.16, 0.2, 0.28, 0.27, 0.29, 0.25, 2.7e+02, 0.035, 0.013, 0.0078 & L00, C03, W10, L10, R10, B14, D16, H17, A13, S09\\
\OID{1342263516} & 1.25, 1.75, 2.50, 3.35, 4.49, 5.73, 7.87, 11.56, 23.67, 71.42, 160.00, 189.00, 890.00, 1300.00 & 0.066, 0.073, 0.068, 0.048, 0.052, 0.056, 0.081, 0.09, 0.083, 0.043, 0.033, 1.5e+02, 0.039, 0.0015 & C03, W10, L10, R10, B14, H17, A13\\
\OID{1342264241} & 1.25, 1.75, 2.50, 3.35, 4.49, 5.73, 7.87, 11.56, 23.67, 71.42, 100.00, 160.00, 189.00, 1300.00 & 0.078, 0.11, 0.12, 0.11, 0.14, 0.13, 0.15, 0.12, 0.1, 0.086, 0.039, 0.25, 1.3e+02, 0.002 & C03, W10, R10, B14, H17, A13\\
\OID{1342265469} & 0.88, 1.25, 1.75, 2.50, 3.35, 4.49, 5.73, 7.87, 11.56, 23.67, 71.42, 160.00, 189.00, 890.00, 1300.00 & 0.0023, 0.034, 0.063, 0.079, 0.073, 0.068, 0.055, 0.049, 0.057, 0.078, 0.12, 0.091, 1.5e+02, 0.0089, 0.0034 & G06, C03, W10, L10, R10, B14, H17, A13\\
\OID{1342265470} & 0.68, 0.88, 1.25, 1.75, 2.50, 3.35, 4.49, 5.73, 7.87, 11.56, 23.67, 70.00, 160.00, 189.00, 890.00, 1300.00 & 2e-05, 0.00019, 0.0014, 0.0023, 0.0043, 0.011, 0.016, 0.017, 0.018, 0.022, 0.073, 0.095, 0.067, 1.5e+02, 0.006, 0.0023 & G07, C03, W10, L10, R10, B14, H17, A13\\
\OID{1342264239} & 1.25, 1.75, 2.50, 3.35, 4.49, 5.73, 7.87, 11.56, 23.67, 100.00, 160.00, 189.00, 860.00, 1300.00, 2600.00 & 0.012, 0.024, 0.037, 0.049, 0.052, 0.045, 0.045, 0.046, 0.053, 0.036, 0.031, 67, 0.00087, 0.0048, 0.0024 & C03, W10, L10, R10, D16, H17, M13, A13, S09\\
\OID{1342263934} & 1.25, 1.75, 2.50, 3.35, 4.49, 5.73, 7.87, 11.56, 23.67, 70.00, 160.00, 189.00, 890.00, 1300.00 & 0.038, 0.049, 0.052, 0.038, 0.039, 0.034, 0.044, 0.051, 0.059, 0.07, 0.044, 1.4e+02, 0.009, 0.0034 & C03, W10, L10, R10, B14, H17, A13\\
\OID{1342264240} & 0.68, 0.88, 1.25, 1.75, 2.50, 3.35, 4.49, 5.73, 7.87, 11.56, 23.67, 70.00, 160.00, 189.00, 850.00, 890.00, 1300.00 & 0.0001, 0.0013, 0.022, 0.04, 0.049, 0.037, 0.043, 0.042, 0.048, 0.048, 0.075, 0.11, 0.15, 1.6e+02, 0.011, 0.0043, 0.002 & M01, L00, C03, W10, L10, R10, B14, H17, K03, R14, A13\\
\OID{1342263489} & 1.25, 1.75, 2.50, 3.40, 4.50, 5.80, 8.00, 12.00, 24.00, 100.00, 160.00, 189.00, 850.00, 890.00 & 0.025, 0.029, 0.024, 0.015, 0.014, 0.013, 0.013, 0.016, 0.055, 0.23, 0.28, 3.2e+02, 47, 0.044 & C03, C13, L08, B14, H17, P16, B11\\
\OID{1342263459} & 1.25, 1.75, 2.50, 3.40, 4.50, 5.80, 8.00, 12.00, 24.00, 189.00 & 0.0073, 0.0088, 0.009, 0.0061, 0.0061, 0.006, 0.0081, 0.013, 0.028, 1.7e+02 & C03, C13, L08, H17\\
\OID{1342263492} & 1.25, 1.75, 2.50, 3.40, 4.50, 5.80, 8.00, 12.00, 24.00, 189.00, 850.00 & 0.067, 0.079, 0.079, 0.052, 0.049, 0.047, 0.044, 0.039, 0.07, 77, 2.2 & C03, C13, L08, H17, P16\\

\enddata
\tablecomments{
    Photometric observations used to create SEDs for model fitting.  A machine readable version of this table is available online.
}
    \tablerefs{
        (B14) \cite{Bulger2014}; (G06) \cite{Guieu2006}; (L10) \cite{Luhman2010}; (M01) \cite{Martin2001}; (R10) \cite{Rebull2010}; (C03) \cite{Cutri2003}; (W10) \cite{Wright2010}; (A13) \cite{Andrews2013}; (M13) \cite{Mohanty2013}; (S09) \cite{Schaefer2009}; (K03) \cite{Klein2003}; (G07) \cite{Guieu2007}; (L00) \cite{Luhman2000}
        (P16) \cite{Pascucci2016}; (H17) This work.
    }
\end{deluxetable}

We combine continuum observations at multiple wavelengths and carry out
continuum radiative transfer (hereafter, RT) modeling to constrain some of the
main disk properties,

with focus on the outer disk radius as it
might affect the
\OI{} emission 
\citep[e.g.][]{Kamp2011}.  Each source's spectral energy distribution (SED) is
shown in Figure \ref{figure-seds} and individual fluxes and references are
given in Table \ref{table-source_photometry} and available online.

\floattable
\begin{deluxetable}{l llllllccc }
\onecolumngrid
\tablecaption{Maximum likelihood parameters for source models.\label{table-model_results}}
\tablehead{
 \colhead{ } &              
 \colhead{$i$} &
 \multicolumn{2}{c}{$R_{\rm in}$} &
 \multicolumn{2}{c}{$R_{\rm out}$} &
 \multicolumn{2}{c}{Log $M_{\rm dust}$} &
 \colhead{$<T_{\rm d}>$} \\
 \cmidrule(lr){3-4}
 \cmidrule(lr){5-6}
 \cmidrule(lr){4-4}
 \colhead{Source} &
 \colhead{($^\circ$)} &
 \multicolumn{2}{c}{(au)} &  
 \multicolumn{2}{c}{(au)} &  
 \multicolumn{2}{c}{($M_\sun$)} &  
 \colhead{(K)}
}
\startdata
    CIDA-1 & {35} & ${0.32}$ & $^{ +1.5}_{ -0.31}$ & ${7.6}$ & $^{ +60}_{ -3.9}$ & ${-4.9}$ & $^{ +0.55}_{ -0.92}$ & {48}  \\ 
FR Tau & {77} & ${0.32}$ & $^{ +1.5}_{ -0.32}$ & ${3.2}$ & $^{ +25}_{ -2.3}$ & ${-6.1}$ & $^{ +0.92}_{ -1.3}$ & {46}  \\ 
FU Tau A & {56} & ${0.00032}$ & $^{ +0.017}_{ -0.00026}$ & ${1.3}$ & $^{ +36}_{ -0.47}$ & ${-7.2}$ & $^{ +1.3}_{ -0.92}$ & 140  \\ 
CFHT-20 & {56} & ${0.0001}$ & $^{ +0.018}_{ -4.4 \times 10^{-5}}$ & ${4.3}$ & $^{ +46}_{ -2.7}$ & ${-5.7}$ & $^{ +0.92}_{ -1.3}$ & {64}  \\ 
J04381486 & {70} & ${0.032}$ & $^{ +0.025}_{ -0.031}$ & ${10}$ & $^{ +5.6}_{ -3.6}$ & ${-4.9}$ & $^{ +0.55}_{ -0.18}$ & {33}  \\ 
GM Tau & {35} & ${0.0032}$ & $^{ +0.015}_{ -0.0031}$ & ${1.8}$ & $^{ +3.2}_{ -0.92}$ & ${-6.8}$ & $^{ +3.1}_{ -0.55}$ & {73}  \\ 
J04393364 & {35} & ${0.01}$ & $^{ +0.55}_{ -0.0094}$ & ${4.3}$ & $^{ +4.6}_{ -3.4}$ & ${-5.7}$ & $^{ +2.4}_{ -1.3}$ & {42}  \\ 
CFHT-4 & {80} & ${0.01}$ & $^{ +0.0078}_{ -0.0099}$ & ${78}$ & $^{ +43}_{ -66}$ & ${-5.3}$ & $^{ +0.18}_{ -0.92}$ & {15}  \\ 
ESO H$\alpha$ 559 & {35} & ${0.032}$ & $^{ +0.53}_{ -0.03}$ & ${33}$ & $^{ +18}_{ -17}$ & ${-4.2}$ & $^{ +1.7}_{ -0.18}$ & {18}  \\ 
Cha H$\alpha$ 1 & {35} & ${0.1}$ & $^{ +0.46}_{ -0.099}$ & ${33}$ & $^{ +88}_{ -24}$ & ${-3.1}$ & $^{ +1.3}_{ -0.55}$ & {16}  \\ 
Hn 13 & {35} & ${0.01}$ & $^{ +0.0078}_{ -0.0099}$ & ${3.2}$ & $^{ +13}_{ -2.3}$ & ${-6.1}$ & $^{ +0.92}_{ -1.3}$ & {64}  \\

\enddata
    \tablecomments{
        The maximum likelihood values are shown for the parameters: $\Rin,\; \Rout\; \mbox{and}\; \Mdust$, along with their corresponding $1\sigma$ confidence intervals.
        $\Tave$ is calculated from the best-fit model selected for the source as described in \S~\ref{sect:individual}.
    }
\end{deluxetable}

We model the SEDs with the 3D axisymmetric Monte Carlo RT code MCMax
\citep{Min2009}.  MCMax self-consistently calculates the disk vertical
structure for a given dust surface density profile and gas-to-dust ratio,
assuming hydrostatic equilibrium and $\Tg=\Td$. The vertical structure of the
dust is calculated from an equilibrium between dust settling and vertical
mixing as described in \cite{Dullemond2004}. Calculation of the disk
temperature and density structure are iterated until a self-consistent solution
is reached. For a given grain size distribution, the only free parameter
controlling the vertical structure is the turbulent mixing strength, which was
not found to be different in modeling the median SEDs of Herbig stars, T~Tauri
stars, and brown dwarfs by \cite{Mulders2012}.

We begin by fitting the stellar parameters and disk inclinations of our sources
using a genetic algorithm.  The stellar luminosity and temperature are taken
from Table~1 and allowed to vary only within their typical uncertainties, 30\%
for the luminosity and $\pm$100\,K for the temperature 
\citep{Luhman2007}.  Higher luminosities
were required for CFHT-4 (2 times higher) and J04381486 (10 times higher) in
order to achieve good fits.  This is likely due to the high inclination of
these disks.  \citet{Ricci2014} estimate a disk inclination of 77$^\circ$
from a resolved millimeter continuum image of CFHT-4, and \citet{Luhman2007}
estimate an inclination of $67-71^\circ$ for J04381486 using a \textit{Hubble
Space Telescope} scattered light observation.  GM Tau required a $\Teff{}$ that was 189\,K lower the the literature value in order to fit the optical/NIR portion of the SED.  These quantities are used to
compute stellar radii from the Stefan-Boltzmann relation.  The source distance
is fixed to that reported in Table~\ref{table-properties}.  The literature
extinction in Table~\ref{table-properties} provides a good fit for all sources.
Eight equally spaced (in cosine) inclinations are sampled to constrain the
viewing angle of the disk.

The general setup for all modeling follows the one described in \citet{Mulders2012} to fit the
median SEDs of TT stars and brown dwarfs.  Given the rather coarse SED sampling at long
wavelengths, and degeneracies in SED modeling \citep[e.g.][]{Woitke2016}, we
limit the number of free parameters. The surface density is a power-law of the
form $\Sigma \propto r^{-1}$ while the minimum and maximum grain sizes are
$a=0.1 \mu$m and 1\,mm with a particle size distribution proportional to
$a^{-3.5}$, following \citet{Mulders2012}.  With this approach, we can explore
the impact of the outer radius on the SED under the explicit assumption that
the surface density slope and grain size distribution are stellar-mass
independent.

To estimate confidence intervals, we run a grid of 4400 models for each source
which explores the disk structure (inner and outer radius) and the disk mass.
For the disk dust mass, 20 logarithmically spaced steps between $10^{-2}$ and
$10^{-9} \Msun$ are sampled.  The disk inner radius is explored using 11
logarithmically spaced steps between $10^{-4}$ and $10^{1}$~au.  For outer
radius, 20 logarithmically spaced steps from 1 to 250~au are used.  These
ranges are chosen such that there is a peak in the Bayesian proability
distribution.  For sources where a clear peak is not observed, physically
unrealistic values justify the boundaries of our grids (e.g., dust masses $>
25$\% stellar mass, $\Rin{}$ $<<$ silicate sublimation radius).

We also determine the likelihood of each model in order to establish confidence
intervals for the parameter space of our model grids.  The likelihood of each
model is calculated by comparing photometric data with model flux densities
using $\mbox{exp}(-\chi^2_R/2)$, where $\chi^2_R$ is the reduced chi-squared
metric.  For photometric detections, we assume a typical uncertainty of 20\%
for each flux measurement.  Upper limits are included in our fitting by using
the modified $\chi^2$ statistic described by \cite{Sawicki2012}.

\begin{figure*}
    \includegraphics[width=\textwidth]{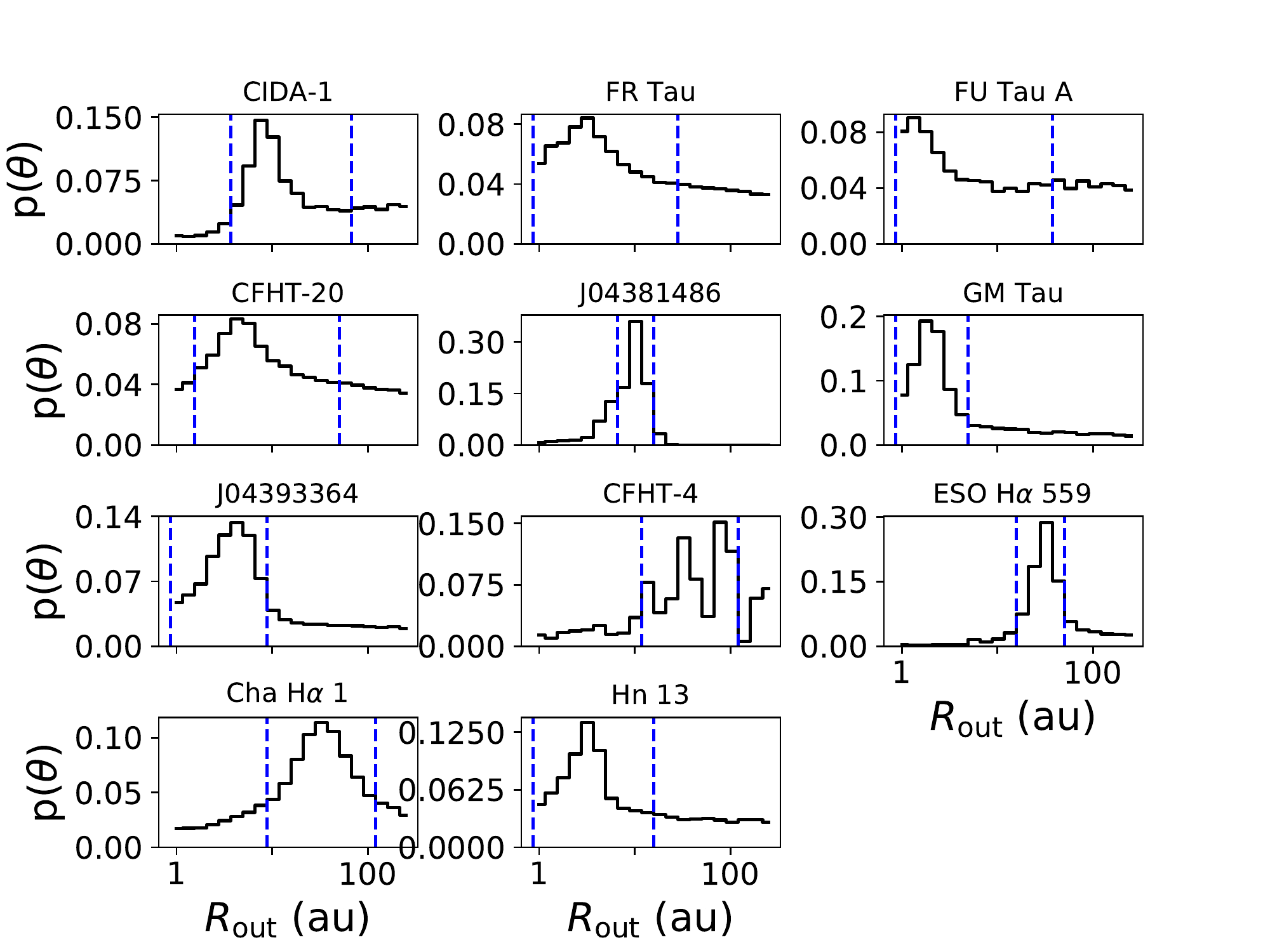}
    \caption{
        \label{figure-Rout_pdfs}
        Bayesian probability distribution of $\Rout$ values.  Vertical blue lines are $1\sigma$ confidence intervals reported in Table \ref{table-model_results}.
    }
\end{figure*}

Relative probabilities are then
calculated for each parameter by summing the likelihood of common parameter
values normalized by the sum of all likelihoods.  68.3\% confidence intervals
for the parameters $\Rout$, $\Rin$ and $\Mdust$ are estimated from Bayesian
probability distributions and reported in Table \ref{table-model_results}.
Figure \ref{figure-Rout_pdfs} shows the probability distribution and
confidence interval for $\Rout$ for each source.  See Appendix \ref{sect:effect_of_outer_radius} for additional information on the confidence intervals and for examples of probability distributions for $\Rin$ and $\Mdust$.

Along with the resulting parameters of our best fit models we also report the mass-averaged disk
temperature ($\Tave$) in Table \ref{table-model_results}.  Here, we take $\Tave$ to be the average of the
dust temperature at each model grid point weighted by the dust mass
at that point. In order to demonstrate the relationship between
this value and the disk outer radius, we choose the model with the maximum
likelihood from the set of models containing the most probable outer radius
value as our fiducial model with which to calculate $\Tave$.

Because we find maximum likelihood values of $R_{\rm out}$ that are
substantially smaller than those of disks around TT stars ($\sim10$ times smaller on
average), we also run a second grid of models keeping $R_{\rm out}$ fixed to
200\,au, a typical value used for modeling TT disks.  These models use the same 20 and 11
steps of $\Mdust$ and $\Rin$, respectively, outlined above.  We use these grids
to similarly derive a $\Tave$ with which to compare our first grid of variable
$\Rout$ against (see Figure \ref{figure-T_vs_L}).

Because heating of the dust in the outer disks of VLMOs by interstellar radiation
may be significant, all of our models include an isotropic interstellar
radiation field (approximated by a diluted 20,000K black body) in combination
with a cosmic microwave background temperature of 2.7\,K to properly compute
the temperature in the outer disk \citep[see][]{Woitke2009}.

\subsection{Outer Radius Parameter Study} \label{sect:methods_grid_models}

In addition to modeling individual sources, we carry out two sets of RT models
to assess the impact of the disk outer radius ($R_{\rm out}$) on the derived
disk mass and $\Tave$.  Both parameter studies are calculated for a range of
stellar masses but in one case $R_{\rm out}$ is held fixed at 200\,au
(hereafter, {\it fixed} $R_{\rm out}$ study) while in the other case
(hereafter, {\it variable} $R_{\rm out}$ study) $R_{\rm out}$ scales linearly
with stellar mass (see also Table \ref{table-parameter_study_model_inputs}).
For both studies, the scaling of disk mass with stellar mass is chosen to
reproduce the best fit relation to the 887,$\micron$ flux densities versus
stellar masses in the Chameleon~I star-forming region \citep{Pascucci2016}.

To cover a large range in stellar mass we combine the evolutionary tracks from
\citet{Baraffe2015} and \citet{Feiden2016} as in \citet{Pascucci2016}. To find
corresponding luminosities and temperatures for each stellar mass we use the
2\,Myr isochrone because this age is a good match to the age of the
star-forming regions we consider in this study. We set 15 equally log-spaced
stellar masses from 0.05~$M_{\odot}$ to 2.239~$M_{\odot}$ which cover from
0.019~$L_{\odot}$ to 6.33~$L_{\odot}$.

All models adopt the same dust properties, turbulent mixing strength, and
surface density power law profiles as those to fit individual SEDs
(\S~\ref{sect:individual}). Inner radii follow a scaling relationship
consistent with the temperature at which silicate sublimation would occur. All
parameters are summarized in Table~\ref{table-parameter_study_model_inputs}.

One important difference between the two parameter studies is the scaling of disk mass with
stellar mass. The {\it fixed} $R_{\rm out}$ study requires a shallower disk
mass-stellar mass relation to reproduce the best-fit 887,$\micron$ flux, and yields a different temperature-luminosity
relation. Implications of this parametric study will be discussed in
\S~\ref{sect:Tdust}.

\begin{deluxetable}{lcc}

\tablecaption{Parameter Study Model Inputs.\label{table-parameter_study_model_inputs}}
\tablewidth{0pt}
\tablehead{
    \colhead{ Parameter}              &
    \colhead{ Variable $R_{\rm out}$} &
    \colhead{ Fixed $R_{\rm out}$}
}
\startdata
    $R_{\rm out}$(au)   &  $200(M_{*}/ M_{\odot})$       &   $200$ \\
    $M_{\rm dust}$      &  $\propto M_*^{2.0}$               & $\propto M_*^{1.2}$ \\
    \hline
    $M_*$ ($M_{\odot}$) &  \multicolumn{2}{c}{ 0.05 - 2.239 } \\
    $R_{\rm in}$(au)    &  \multicolumn{2}{c}{ $0.05 \times L_*^{0.5}$ }
\enddata
\end{deluxetable}

\section{Results}\label{sect:results}

\begin{figure}
\centering
\includegraphics[width=0.5\textwidth]{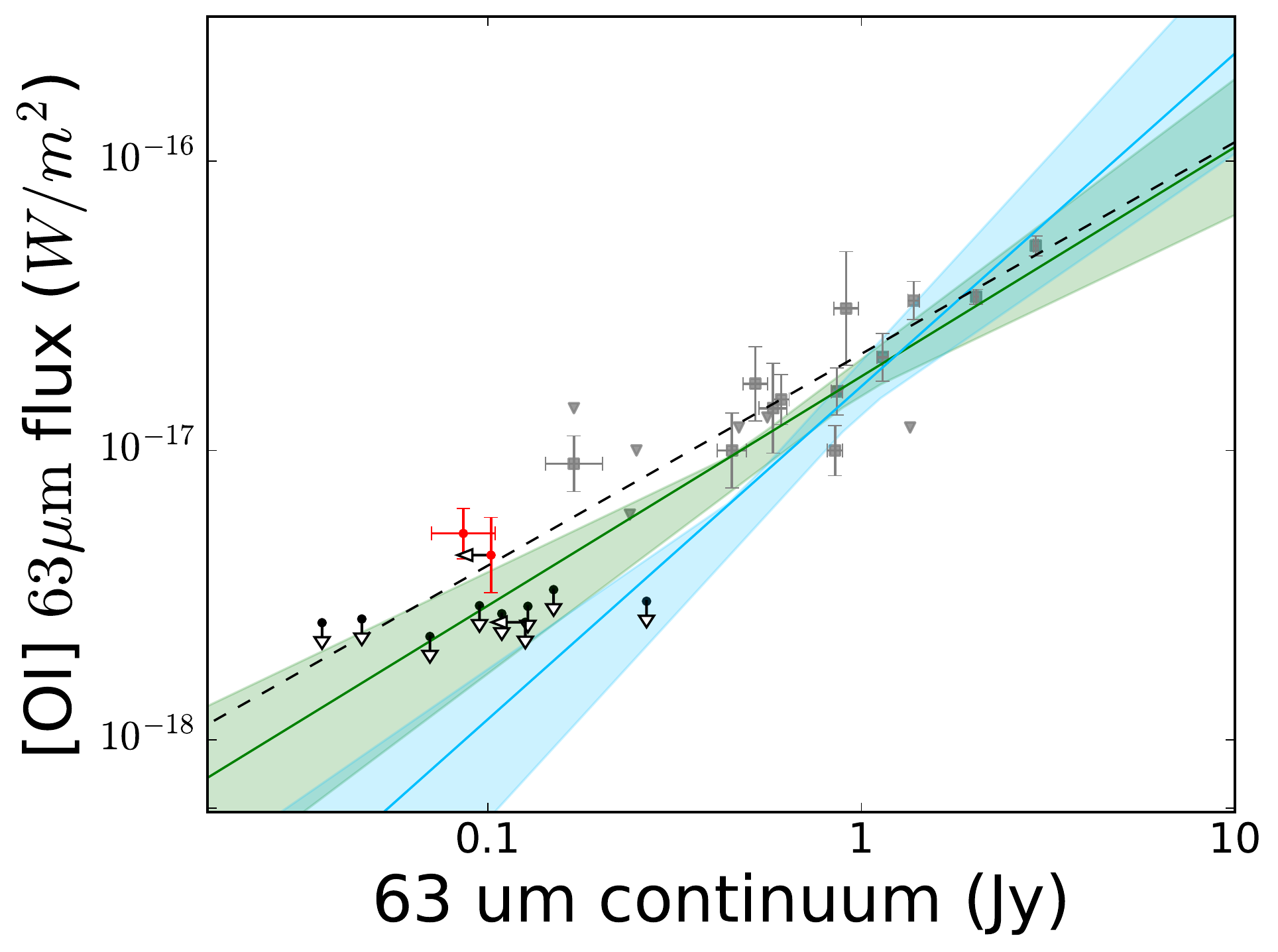}
\caption{
    \label{figure-oi_vs_continuum}
    Fluxes are scaled to a distance of 140pc as in \cite{Howard2013}. Red
    points are [OI] detections from this work.  Black points are this work
    with arrows indicating upper limits in continuum and/or line flux
    observations.  Grey points are full disks from \citeauthor{Howard2013} 
    (squares are detections and triangles are upper limits), with the dashed
    line showing their best fit.  Green line is a refitting of
    the \citeauthor{Howard2013} data using their upper limits.  Blue line is the fit including
    our sources.  Shaded areas are $1\sigma$ confidence intervals.
}
\end{figure}

\subsection{Faint \OI{} emission from very low-mass star and brown dwarf disks}\label{sect:faint}

As mentioned in \S~\ref{sect:oidet} we detect the \OI{} line
only toward 2 out of the 11 VLMO disks in the sample. Figure
\ref{figure-oi_vs_continuum} shows the [OI] line flux versus the
continuum of our sample (red and black symbols) in the context of literature
values for more luminous/massive objects (grey symbols). The non-detections are
somewhat surprising given that most sources should have been detected if they
followed the line luminosity-continuum relation for TT stars with no known jets from
\cite{Howard2013} (dashed line). This suggests that VLMO
disks may be under-luminous in the \OI{} line. 

In order to consider the possibility that the trend in the line
flux-continuum relation does not extend to the VLMO regime
and that the VLMO \OI{} emission is under-luminous, we have re-fit the same data and
included the TT sources with upper limits which Howard et al. omitted from their fit.
By doing this we are able to then re-fit the full TT sample (detections and
upper limits) with our VLMO non-detections
\footnote{We consider only non-detections from our sample.  Of the two VLMOs with detected [OI] lines, one is excluded from our analysis because it has a known jet, and the other is excluded because it has only an upper limit for its $63,\micron$ continuum.}
(see Figure
\ref{figure-oi_vs_continuum}).

To do this we use the Bayesian method of regression fitting described in 
\cite{Kelly2007}\footnote{Implemented by Josh Meyers in Python (Jan 16 2016 commit).} which takes
into account non-detections and errors bars.

We exclude FU~Tau~A from the fitting
because it is a known outflow source \cite{Monin2013}. Table
\ref{table-fitting} gives the fit coefficients.

\begin{deluxetable}{l cc l}
    \tablecaption{[OI] vs 70um Continuum Fitting Results\label{table-fitting}}
\tablehead{
    \colhead{} &
    \multicolumn{2}{c}{Linear Regresion} \\
    \cmidrule(lr){2-3}
    \colhead{Description}  &
    \colhead{Intercept} &
    \colhead{Slope}
}
\startdata
    TT Disks  & $0.89 \pm 0.18$ & $-16.74 \pm 0.06 $ \\
    TT + VLMO & $1.14 \pm 0.22$ & $-16.78 \pm 0.09 $ \\
\enddata

\end{deluxetable}

The blue line in Figure \ref{figure-oi_vs_continuum} shows the best fit with
the addition of our sources.  VLMO disks may be under luminous in the \OI{} line,
as suggested by the brightest (in continuum) half of the VLMO sample with no [OI]
detection, but the upper limits are not stringent enough to conclude that they
are actually under luminous at a confidence greater than 89\%.  On average our
samples appear to be under-luminous by a factor of 1.8.

\begin{figure*}[tbh]
\makebox[\textwidth]{%
\includegraphics[width=0.55\textwidth]{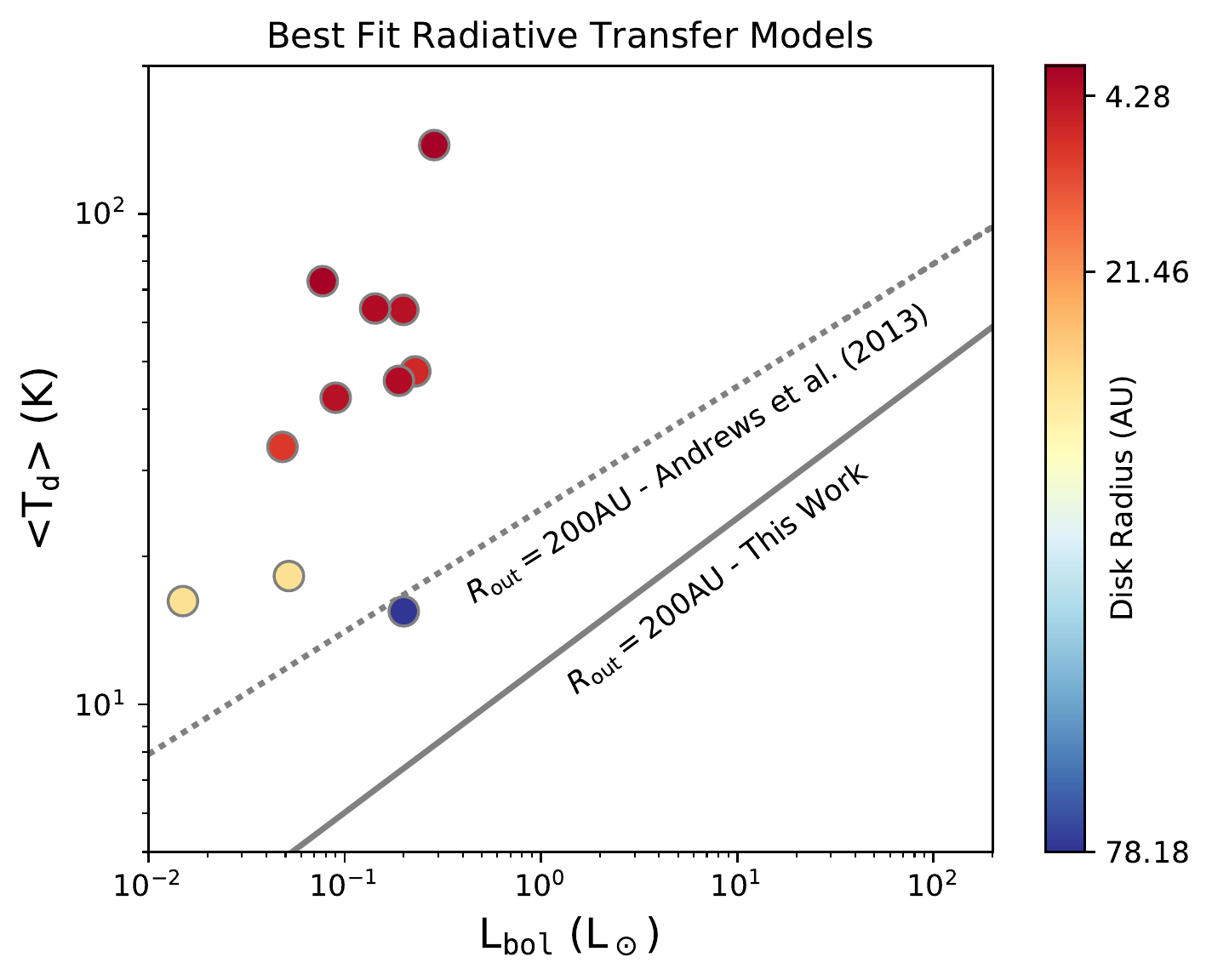}%
\hfill    
\includegraphics[width=0.55\textwidth]{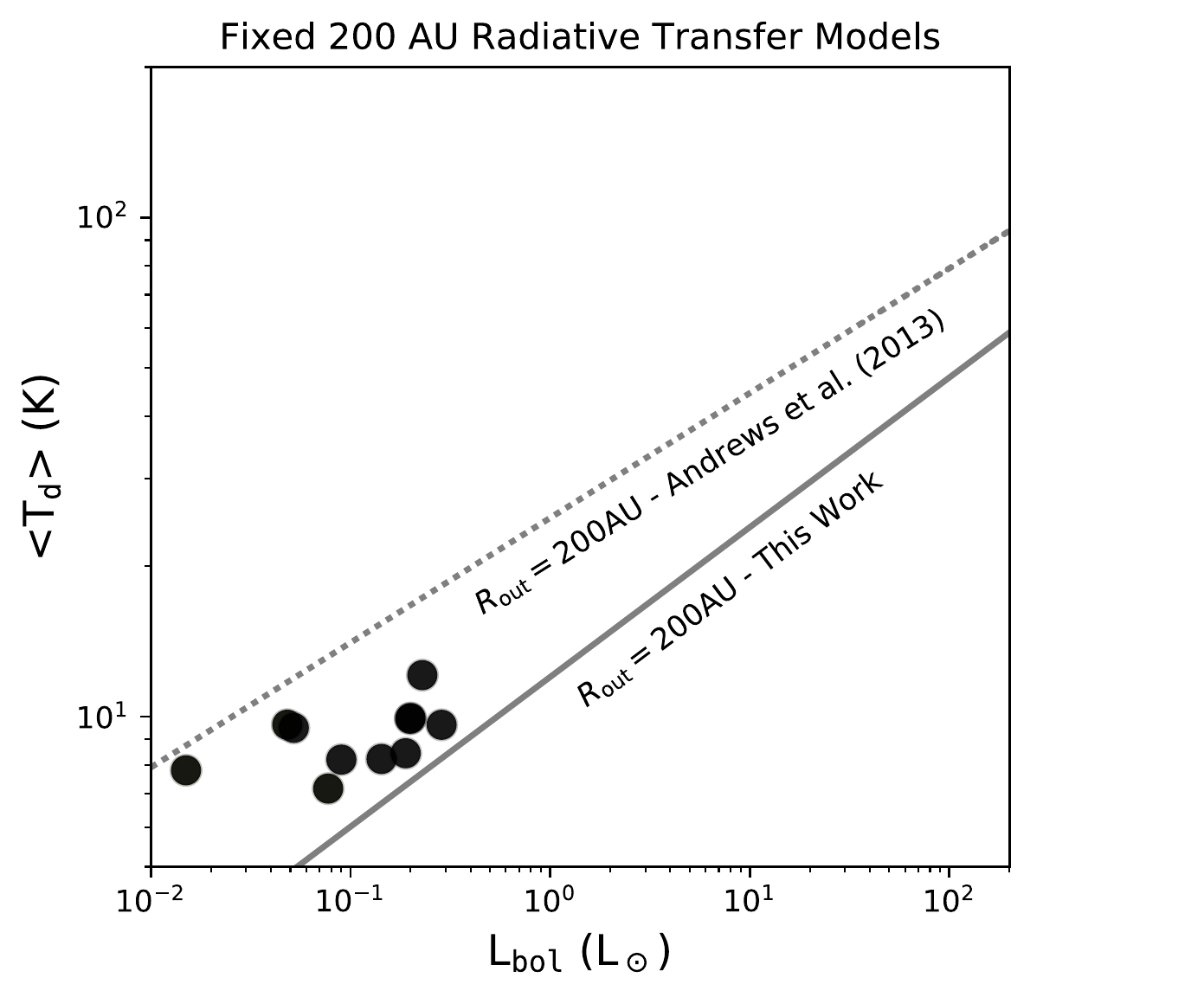}%
}

\caption{
    Circles are outputs from the RT modeling of our \Herschel 
    sources.  The left panel shows the best fit of these models where the outer disk
    radius is a free parameter.  The right panel shows the best fits where disk
    radius is fixed at 200~au.  Both panels show the best-fit temperature-luminosity relation of our
    fixed radius parameter study models (lower solid line) and the \citet{Andrews2013} fit (dotted line).
}
\label{figure-T_vs_L}
\end{figure*}

\subsection{Small dust disks}\label{sect:small}

Our RT modeling suggests greater likelihoods for outer disk radii that are
smaller (maximum likelihoods $\leq 78$~au with a mean value of 15~au) than typical values ($>
150$ au) for disks around T-Tauri stars
\cite[e.g.][]{Isella2009,Andrews2009,Guilloteau2011}.  We also note that for
every source, the confidence intervals (see Figure \ref{figure-Rout_pdfs}) lie
entirely well below 200~au.  In fact, with the exception of CFHT-4, the
confidence intervals lie entirely below 100~au, and the Bayesian probability
distributions decrease towards larger outer radii.

Although these best fit models may not be unique solutions and stellar-mass dependencies in other disk parameters might yield equally good fits \citep[e.g.][]{Woitke2016}, this approach shows that smaller disks around sub-stellar objects are consistent with the SEDs.

Disk sizes are mostly unknown in the brown dwarf regime.  For our sample
spatially resolved millimeter images exist only for CIDA-1, CFHT-4, and
ESO~H$\alpha$~559 \citep{Ricci2014,Pascucci2016}. Detailed modeling has been
carried out for the first two with estimated disk radii of $66\pm^{14}_{12}$~au
for CIDA-1 and greater than 80~au for CFHT-4 \citep{Ricci2014}.
ESO~H$\alpha$~559 was resolved at $887,\micron$ and a fitted elliptical
Gaussian resulted in a FWHM of 0.23\arcsec x 0.15\arcsec\ (37~au x 24~au,
\cite{Pascucci2016}).  The maximum likelihood outer radius values found by our
modeling of CFHT-4 (78~au) and ESO~H$\alpha$~559 (33~au) are in good agreement
with these findings.

\cite{Ricci2014} find an outer radius for CIDA-1 which differs significantly
from our value of 7.6~au, however, our modeling does not reject large sizes and
the $1\sigma$ confidence interval of 60~au is consistent with the estimate of
Ricci et al.  These three disks are all larger than the median VLMO disks in
our sample but they are also three of the four brightest disks.  This suggests
that smaller disks are indeed more common in the brown dwarf regime (see also
\cite{Testi2016}).  Additionally, \cite{Luhman2007} found that their models of
J04381486 with outer radii of 20 and 40~au agreed ``reasonably well'' with
\Hubble{} Wide Field Planetary Camera observations.  These values are larger
than our finding of $\Rout{}={10}^{ +5.6}_{-3.6}$~au, in line with observations
showing that sub-micron grains and gas disks, as probed via scattered light
images, are typically larger than dust disks traced at millimeter wavelengths
\citep[e.g.][]{deGregorio-Monsalvo2013}.

If VLMO disks are indeed smaller than those around TTs, an
important implication is that they are hotter than 
we would estimate if they were the same size as disks around TTs,
i.e. their $\Tave$ is higher.

The left panel of Figure~\ref{figure-T_vs_L}
illustrates this point: circles are our best fit values color-coded by
disk outer radius while the dotted and solid lines are the $\TvL$ relations by
\citet{Andrews2013} and by our {\it fixed} $R_{\rm out}$ models
(\S~\ref{sect:methods_grid_models}).  In both models the dust disk radius
is 200\,au regardless of stellar luminosity/mass but our {\it fixed} $R_{\rm
out}$ models are significantly cooler because we compute the disk vertical
structure self-consistently which results in less vertically extended disks.
Regardless of these differences, it is clear that our best-fit RT models point
to disk radii smaller than 200\,au and higher $\Tave$ for VLMO
disks than typically assumed. The SED fits with fixed $R_{\rm out}$ have
temperatures in between those predicted by the \citet{Andrews2013} and our
{\it fixed} $R_{\rm out}$ grid relations (right panel of
Figure~\ref{figure-T_vs_L}).

Note that the \textit{Herschel} models with fixed 200~au outer radius fall above
(and not on) the trend-line produced by our 200~au parameter models for two
reasons. First, the Herschel sources were selected to be brighter, and hence
hotter, disks whereas the model grid was fit to the median disk mass in
Chameleon. Second, the parameter models use disk masses and inner disk radii
that scale with stellar mass and bolometric luminosity respectively, whereas
our best-fit SED models allow these value to vary for a best fit with
photometry.

Interestingly, our variable $\Rout$ grid VLMO models are found to be optically
thick at $63,\micron$ for all of our sources with the exception of CFHT-4 (the
source with the largest $\Rout$) and at $850,\micron$ for all but two (CFHT-4
and FR~Tau).  This is different than what is proposed in Harvey et al. who,
however, targeted a fainter $70,\micron$ sample of brown dwarf disks. Measuring dust disk sizes
will be important to establish if, and by how much, brown dwarf disks are optically
thick at these wavelengths.

\subsection{Disk Temperature-Stellar Luminosity Relation}\label{sect:Tdust}

A typical approach to measure dust disk masses is to obtain a sub-mm/mm
data point, a source distance, assume optically thin emission and dust opacity,
and an average dust temperature ($\Tave$) to be used in the Planck
function \citep[e.g.][]{Beckwith1990}. This same approach has been recently
used to estimate dust disk masses for hundreds of stars in nearby star-forming
regions and associations. As discussed in \citet{Pascucci2016} such
estimates, as well as the disk mass-stellar mass scaling relation, are very
sensitive to the assumed $\Tave$ and how it scales with stellar
luminosity.

\begin{table*}
    \caption{Effect of model assumptions on total disk mass (gas + dust) estimates\label{tab:disk_masses}}
\centering

\begin{tabular}{l|ccc|ccc|ccc}
    \hline\hline
    &
    \multicolumn{6}{c|}{Fixed Radius Estimates} &
    \multicolumn{3}{c}{Scaled Radius Estimates} 
    \\
    &
    \multicolumn{3}{c|}{This work} &
    \multicolumn{3}{c|}{Andrews et al. (2013)} &
    \multicolumn{3}{c}{This work} 
    \\
    &
    \multicolumn{2}{c}{Disk Mass} &
    \multicolumn{1}{c|}{$\Tave$} &
    \multicolumn{2}{c}{Disk Mass} &
    \multicolumn{1}{c|}{$\Tave$} &
    \multicolumn{2}{c}{Disk Mass} &
    \multicolumn{1}{c}{$\Tave$}
    \\
    &
    \multicolumn{1}{c}{($\Msun$)} &
    \multicolumn{1}{c}{($\Mstar$)} &
    \multicolumn{1}{c|}{(K)} &
    \multicolumn{1}{c}{($\Msun$)} &
    \multicolumn{1}{c}{($\Mstar$)} &
    \multicolumn{1}{c|}{(K)} &
    \multicolumn{1}{c}{($\Msun$)} &
    \multicolumn{1}{c}{($\Mstar$)} &
    \multicolumn{1}{c}{(K)}
    \\
    \hline
    \OID{1342268646} & 0.017 & 0.09 & 7.40 & 0.0036 & 0.019 & 16.71& 0.0040 & 0.021 & 15.67 \\
    \OID{1342265470} & 0.274 & 5.05 & 2.45 & 0.0039 & 0.071 & 6.65& 0.0005 & 0.008 & 20.29 \\
    \OID{1342264239} & 0.008 & 0.11 & 3.25 & 0.0003 & 0.004 & 8.42& 0.0001 & 0.001 & 18.99 \\
    \OID{1342264240} & 0.004 & 0.03 & 6.01 & 0.0006 & 0.005 & 14.06& 0.0004 & 0.004 & 16.45 \\
    \OID{1342263489} & 0.069 & 0.48 & 4.96 & 0.0078 & 0.055 & 11.97& 0.0043 & 0.030 & 17.21 \\
    \hline
\end{tabular}
\end{table*}

\begin{figure}
\includegraphics[width=\linewidth]{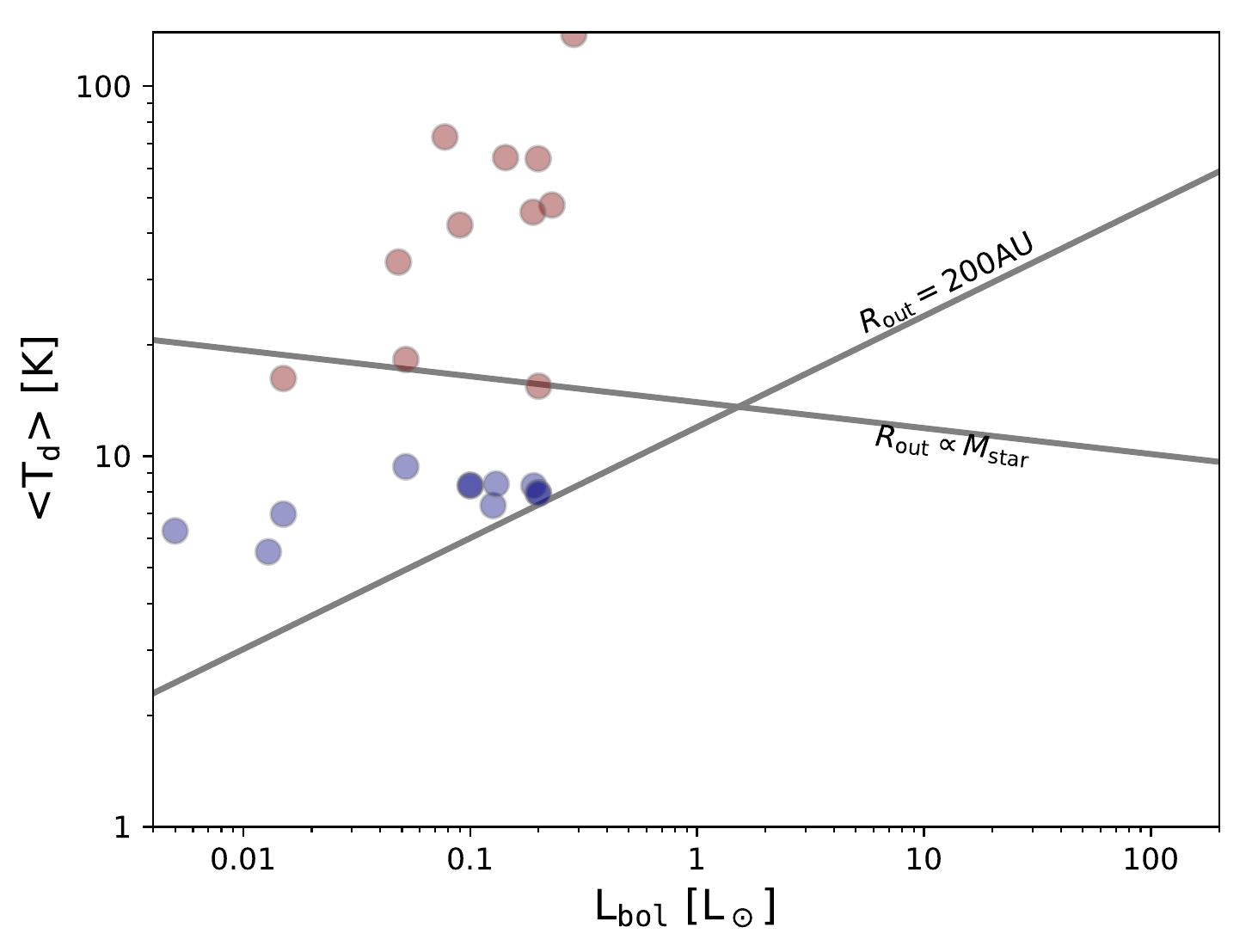}
\caption{
    \label{figure-T_vs_L_3}
    $R_{out} \propto M_{star}$ shows the best fit of our grid of RT
    models where the disk's outer radius is scaled to the stellar
    mass.  $R_{out} = 200$~au shows the best fit of our grid of RT
    models where the disk's outer radius is fixed to 200~au (it is the
    same line seen in the right and left panels of Figure \ref{figure-T_vs_L}).  Red
    dots are our best fit models (figure \ref{figure-T_vs_L} left panel) and
    the blue dots are the {\it fixed} $\Rout$ grid (200~au) models (Figure
    \ref{figure-T_vs_L} right panel).
}
\end{figure}

In the previous section we showed that the SEDs of VLMOs can be well reproduced
    with disks that have smaller radii than TT disks and, as such, do not fall
    on the temperature-luminosity relation derived by \citet{Andrews2013}. Instead, they
    are hotter, plotting above this relation (see Figure~\ref{figure-T_vs_L_3} and
    Table~\ref{table-model_results}).  We use the two RT grids discussed in
    \S~\ref{sect:methods_grid_models} to quantify the difference in the $\TvL$
    relation for fixed outer radii disks and radii scaling with stellar mass
    (see Figure \ref{figure-T_vs_L_3}).  The {\it fixed} $R_{\rm out}$ models
    result in mass-averaged disk temperatures that decrease with stellar
    luminosity and become lower than 10\,K for brown dwarfs.  This is somewhat
    surprising given that dust grains in giant molecular clouds, heated by the
    interstellar radiation field, stabilizes at $\sim10$\,K \citep{Mathis1983}.
    We remind the reader that we have included interstellar radiation in our
    modeling (see \S~\ref{sect:individual}) but, as also found in
    \citet{vanderPlas2016}, this extra heating does not change appreciably the
    dust disk temperature (see in particular their Figure~5d).

In the very different assumption of outer radii scaling with stellar mass,
$\Tave$ remains confined within $\sim$10-20\,K over four orders of magnitude in
luminosity and shows an opposite behaviour by slightly increasing toward the
lower-luminosity/lower-mass objects.  For example, the difference in $\Tave$
for the two assumptions when $L_* = 0.1 L_{\odot}$ is 10.4~K which results in
mass estimates that differ by a factor of 9.

\section{Discussion}\label{sect:discussion}

In this paper we start from the zero-order assumptions that disk geometry and
dust opacity are stellar-mass independent.  With these assumptions, our RT
modeling of each source suggests that dust disk outer radii range from 1.3 to
78~au.  This is significantly less than the 200~au radius typically assumed for
TT disks. Although small disk radii may not be a unique solution, they agree
well with the few previously resolved VLMO disks.  As such, it becomes necessary
to consider the effect of smaller disk sizes when calculating their masses and
interpreting the \OI{} line emission.

\subsection{Mass estimates for small disks}\label{sect:discussion-small}

Disk mass estimates using a $\Tave$ taken from a $L_{\rm bol}$
relationship with a fixed outer radius will ultimately result in temperatures
that are too low for low luminosity objects and temperatures that are too high
for high luminosity objects (Fig. \ref{figure-T_vs_L_3}).  Consequently this
will result in over-predicting and under-predicting the dust disk masses of low
and high luminosity sources respectively.

$\Tave$ can vary by a factor of 3 to 5 times for luminosities between 0.01 and
100 $L_\odot$.  If we consider VLMOs to be represented by $\Lbol \in [0.005, 0.2]
\Lsol$ (The range of our Herschel sample) we find $\Tave$ to be lower by a
factor of 8 to 2 with a mean difference (in logarithmic space) of 5 times.  The
disagreement in the two treatments is much less for TT objects where
considering typical $\Lbol \in [1.5, 8] \Lsol$ (comparable to the TT disks
considered by Howard et al 2013) results in a over-estimate of $\Tave$ by a
factor $\sim1.5$. 
In addition, the fixed outer radius relationship

gives too low of a $\Tave$ in the VLMO regime.
Luminosities below
$0.053 \Lsun$ (which would include 4 of our 11 sources) result in $\Tave$
under 5~K.

Ultimately these disagreements in $\Tave$ are significant because they result
in an even greater discrepancy in mass predictions.  For example, using the
fixed $\TvsL$ relationships and applying them to the source J04381486 ($0.005
\Lsun$; $F_{890,\micron}=6$~mJy), a total disk mass (assuming a gas-to-dust
radio of 100) of $5$ and $0.07\Mstar{}$ is estimated (depending on whether the
vertical structure is self consistently calculated or not) due to the low
$\Tave$ of 6.7~K and 2.5~K respectively. Table \ref{tab:disk_masses} compares
the effects of model assumption on estimated disk masses for the 5 sources in
our sample with $\sim 887,\micron$ detections.

Because the dust disk size likely depends on many factors including disk mass,
height, turbulent mixing, age, grain-growth and drift rates, perhaps even
more so than it depends upon host luminosity, characterizing the dust-disk
outer radius becomes critical in understanding disk masses.

\subsection{[OI] emission in small disks}

For our 11 observed sources we had only 2 \OI{} line detections.  Figure
\ref{figure-oi_vs_continuum} shows the fit (dashed line) to TT disks found by
\citet{Howard2013} which we used to set the sensitivity of our observations.
Based on this we expected more detections but over two-thirds of our sources
resulted in upper limits which fall below their fit.  

\begin{figure}
\begin{center}
\includegraphics[width=\linewidth]{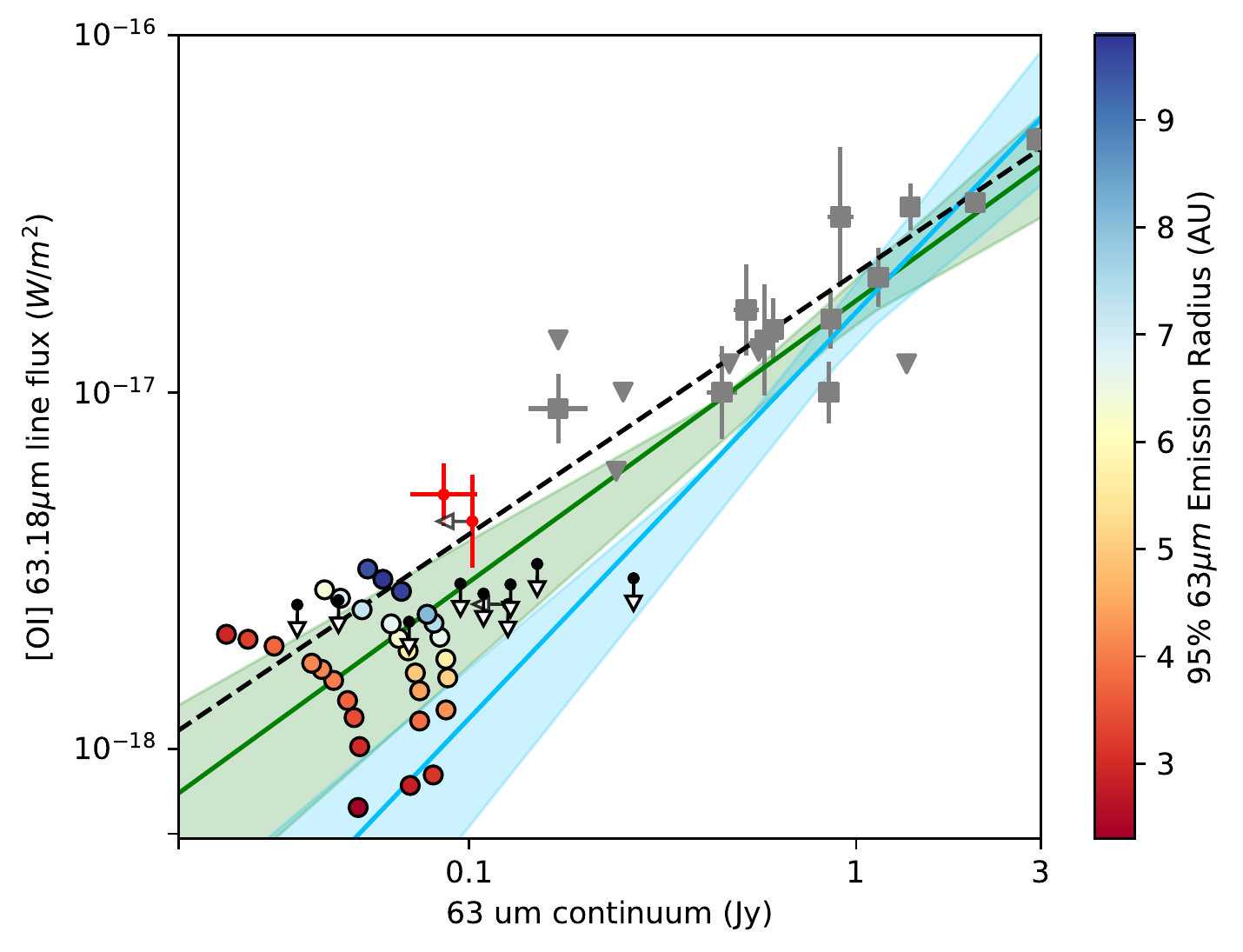}
\caption{
    Line and continuum emission from thermochemical models by
    \cite{Greenwood2017} (color-filled circles) superimposed on data and fits
    described in Figure \ref{figure-oi_vs_continuum}.  Isarithms of constant
    mass (1, 4, and 8 $\times10^{-4} \Msun$) and flaring index show how smaller
    disks lead to a rapid decrease in [OI] line flux.
}
\label{figure-thermochemical-oi}
\end{center}
\end{figure}

In order to assess how feasible it is that these disks are under-luminous (and
to pinpoint a likely origin of this under-luminosity), we compare our
observations to the thermochemical models by \citet{Greenwood2017}. These
models use 2D RT and a complex chemical network on top of a
parametrized disk structure, allowing investigations into the effects of disk
geometry on the \OI{} line flux by computing a grid of models of varying disk
masses and radii.

Figure \ref{figure-thermochemical-oi} shows that for a given mass, a decreasing
of the disk size leads to a decrease in [OI] emission while continuum emission
increases or stays constant. This is consistent with the [OI] underluminosity
of VLMOs being caused by smaller disk sizes.

\begin{figure}
\begin{center}
\includegraphics[width=\linewidth]{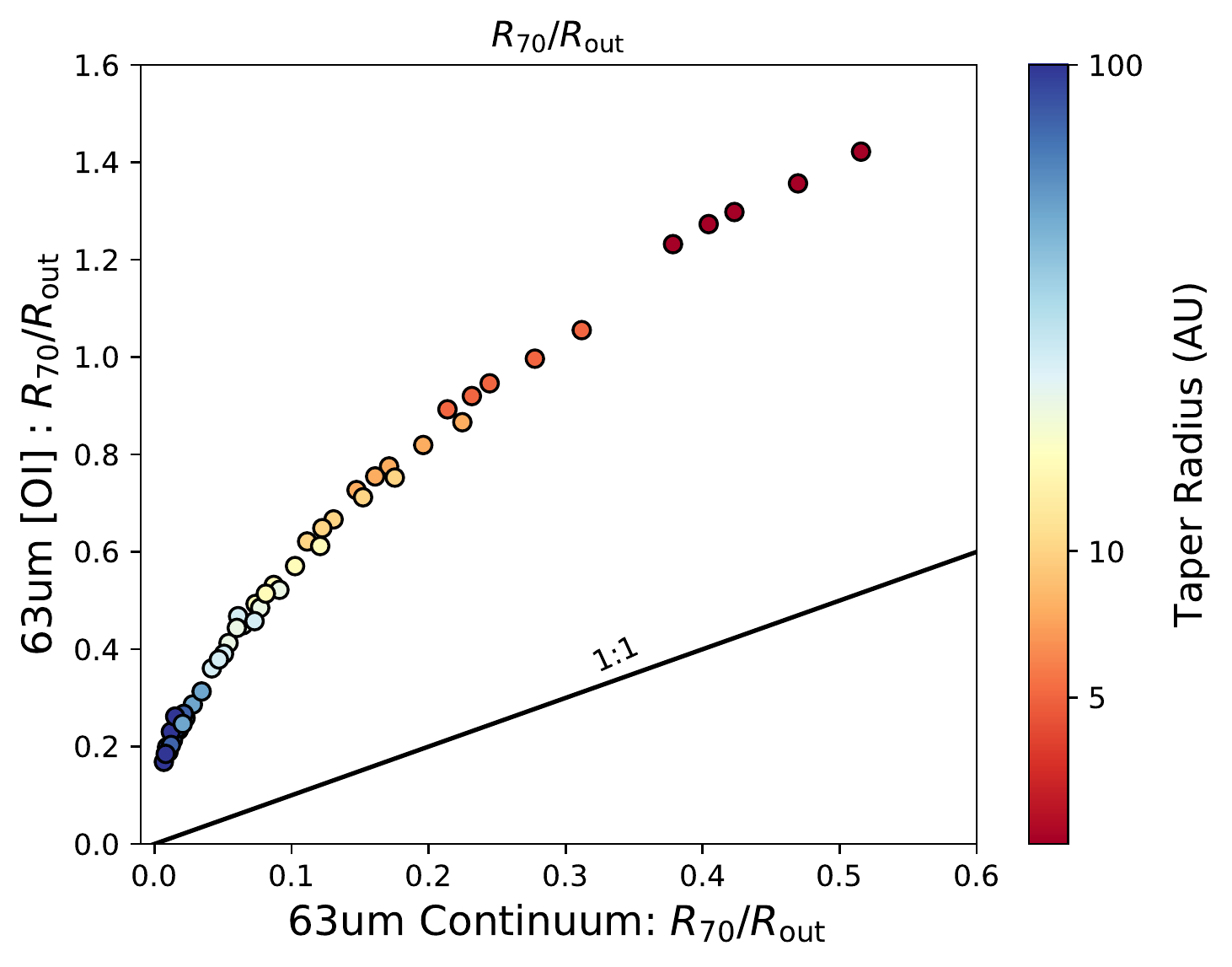}
    \caption{
        Line and continuum emission from thermochemical models by \cite{Greenwood2017}
        showing that 70\% of \OI{} emission comes from
        larger relative distances as the disk size decreases.
        70\% of the given emission ([OI] or continuum) originates from within the R$_{70}$ radius.
    }
\label{figure-R70}
\end{center}
\end{figure}

The reason for the reduced [OI] emission is due to the radii at which the
emission originates.  Figure \ref{figure-R70} shows the radius within which
70\% of the [OI] and continuum emissions originate normalised to the taper
radius\footnote{Taper radius is the location of an exponential tapering-off of
the disk density power law.  $R_{\rm taper} = R_{\rm out} / 8$.} , a measure of
the disk size. For all disk sizes, the [OI] emission originates from larger
radii than the continuum.  For large disks, the majority of the emission is
well within the taper radius for both the \OI{} line flux and the continuum
flux density.  However, as the disks become smaller (going from blue to red in
the figure), the area of \OI{} line emission falls beyond the taper radius
while the continuum emission remains well within.  Thus, one possible
contribution to the observed under-luminosity is that the line-emitting area of
\OI{} is disproportionately small in VLMO disks.

Another contributing factor may be that VLMO disks tend to be less flared than
their higher-mass counterparts, for which there is some observational evidence
\citep[however see \citealt{Harvey2012}]{SzHucs2010,Liu2015}.  The models by
\citet{Greenwood2017} show that changing the flaring index ($\beta$) from 1.2 to 1.15,
1.10, and 1.05 decreases the \OI{} emission in a VLMO disk to 73\% and 40\%, and
13\% of the $\beta=1.2$ levels respectively, while the continuum emission
decreases to 87\%, 40\%, and 19\% of the $\beta=1.2$ levels (Greenwood,
personal communication). These results suggest that the \OI{} line flux is more
sensitive to disk flaring than the continuum flux, causing a sample of
weakly-flared disks to be under-luminous in their \OI{} emission.

\section{Summary and Conclusions}

Starting from a sample of 11 luminosity-selected very low-mass stars and brown
dwarfs with disks, we use the PACS spectrometer aboard the \HSO to measure the
\OI{} line fluxes and model the spectral energy distributions.  The analysis of
10 out of our 11 sources is consistent with the previous analysis by
\cite{Riviere-Marichalar2016}, however using a newer data reduction pipeline,
we report a new [OI] 63um detection toward FU~Tau~A.

Our main findings are:

\begin{itemize}

    \item We detect only two sources in \OI{} despite a 3 times better sensitivity than for T Tauri disks, 
    suggesting that disks around very low-mass star and brown dwarf disks are underluminous in \OI{}. 

    \item Assuming that the disk geometry and opacity are stellar-mass
        independent, we find that disk models with outer radii in the range of
        1.3--78~au have the highest maximum likelihood values when compared
        to SEDs. These radii are significantly smaller than those of T~Tauri
        star disks (22--440~au).

    \item If very low-mass star and brown dwarf disks are indeed smaller and hotter than T Tauri disks they do not
        follow previously derived temperature-luminosity relationships.  A disk
        outer radius that scales linearly with stellar mass results in almost
        flat $\TvsL$ relationship. This results in higher average temperatures
        for VLMO disks and smaller disk mass estimates.

    \item Using thermochemical models we find that smaller disks result in 
        lower \OI{} fluxes which may explain the non-detections in the sample. 

\end{itemize}

A smaller size of protoplanetary disks around sub-stellar objects has a large
impact on the derived disk masses from large surveys of star-forming regions.
High-resolution ALMA observations are necessary to quantifying how the disk
radius scales with stellar mass to correctly gauge the planet-forming potential
of very low-mass stars and brown dwarfs.

\begin{acknowledgements}

The Herschel spacecraft was designed, built, tested, and launched under a
contract to ESA managed by the Herschel/Planck Project team by an
industrial consortium under the overall responsibility of the prime
contractor Thales Alenia Space (Cannes), and including Astrium
(Friedrichshafen) responsible for the payload module and for system testing
at spacecraft level, Thales Alenia Space (Turin) responsible for the
service module, and Astrium (Toulouse) responsible for the telescope, with
in excess of a hundred subcontractors.
HIPE is a joint development by the
Herschel Science Ground Segment Consortium, consisting of ESA, the NASA
Herschel Science Center, and the HIFI, PACS and SPIRE consortia.
An allocation of computer time from the UA Research Computing High Performance
Computing (HPC) and High Throughput Computing (HTC) at the University of
Arizona is gratefully acknowledged.  This material is based upon work supported
by the National Science Foundation under Grant No. 1228509.
Support for this work, part of the NASA Herschel Science Center Theoretical Research/Laboratory Astrophysics Program, was provided by NASA through a contract issued by the Jet Propulsion Laboratory, California Institute of Technology under a contract with NASA (Grant No. 1483963).

This research has made use of the SIMBAD database and the VizieR catalogue
access tool, CDS, Strasbourg, France.

{\it Facilities:} \facility{Herschel (PACS)}

\end{acknowledgements}

\bibliography{bd-disks-references,bd-disks-custom-references}

\begin{thebibliography}{}
\expandafter\ifx\csname natexlab\endcsname\relax\def\natexlab#1{#1}\fi

\bibitem[{{Ahmic} {et~al.}(2007){Ahmic}, {Jayawardhana}, {Brandeker}, {Scholz},
  {van Kerkwijk}, {Delgado-Donate}, \& {Froebrich}}]{Ahmic2007}
{Ahmic}, M., {Jayawardhana}, R., {Brandeker}, A., {et~al.} 2007, \apj, 671,
  2074

\bibitem[{{Andrae}(2010)}]{Andrae2010}
{Andrae}, R. 2010, ArXiv e-prints, arXiv:1009.2755

\bibitem[{{Andrews} {et~al.}(2013){Andrews}, {Rosenfeld}, {Kraus}, \&
  {Wilner}}]{Andrews2013}
{Andrews}, S.~M., {Rosenfeld}, K.~A., {Kraus}, A.~L., \& {Wilner}, D.~J. 2013,
  \apj, 771, 129

\bibitem[{{Andrews} {et~al.}(2009){Andrews}, {Wilner}, {Hughes}, {Qi}, \&
  {Dullemond}}]{Andrews2009}
{Andrews}, S.~M., {Wilner}, D.~J., {Hughes}, A.~M., {Qi}, C., \& {Dullemond},
  C.~P. 2009, \apj, 700, 1502

\bibitem[{{Andrews} {et~al.}(2010){Andrews}, {Wilner}, {Hughes}, {Qi}, \&
  {Dullemond}}]{Andrews2010}
---. 2010, \apj, 723, 1241

\bibitem[{{Apai} {et~al.}(2005){Apai}, {Pascucci}, {Bouwman}, {Natta},
  {Henning}, \& {Dullemond}}]{Apai2005}
{Apai}, D., {Pascucci}, I., {Bouwman}, J., {et~al.} 2005, Science, 310, 834

\bibitem[{{Baraffe} {et~al.}(1998){Baraffe}, {Chabrier}, {Allard}, \&
  {Hauschildt}}]{Baraffe1998}
{Baraffe}, I., {Chabrier}, G., {Allard}, F., \& {Hauschildt}, P.~H. 1998, \aap,
  337, 403

\bibitem[{{Baraffe} {et~al.}(2015){Baraffe}, {Homeier}, {Allard}, \&
  {Chabrier}}]{Baraffe2015}
{Baraffe}, I., {Homeier}, D., {Allard}, F., \& {Chabrier}, G. 2015, \aap, 577,
  A42

\bibitem[{Barlow(1989)}]{Barlow1989}
Barlow, R.~J. 1989, Statistics: a guide to the use of statistical methods in
  the physical sciences, Vol.~29 (John Wiley \& Sons)

\bibitem[{{Bate}(2012)}]{Bate2012}
{Bate}, M.~R. 2012, \mnras, 419, 3115

\bibitem[{{Beckwith} {et~al.}(1990){Beckwith}, {Sargent}, {Chini}, \&
  {Guesten}}]{Beckwith1990}
{Beckwith}, S.~V.~W., {Sargent}, A.~I., {Chini}, R.~S., \& {Guesten}, R. 1990,
  \aj, 99, 924

\bibitem[{{Billot} {et~al.}(2012){Billot}, {Morales-Calder{\'o}n}, {Stauffer},
  {Megeath}, \& {Whitney}}]{Billot2012}
{Billot}, N., {Morales-Calder{\'o}n}, M., {Stauffer}, J.~R., {Megeath}, S.~T.,
  \& {Whitney}, B. 2012, \apjl, 753, L35

\bibitem[{{Bulger} {et~al.}(2014){Bulger}, {Patience}, {Ward-Duong}, {Pinte},
  {Bouy}, {M{\'e}nard}, \& {Monin}}]{Bulger2014}
{Bulger}, J., {Patience}, J., {Ward-Duong}, K., {et~al.} 2014, \aap, 570, A29

\bibitem[{{Cutri} {et~al.}(2003){Cutri}, {Skrutskie}, {van Dyk}, {Beichman},
  {Carpenter}, {Chester}, {Cambresy}, {Evans}, {Fowler}, {Gizis}, {Howard},
  {Huchra}, {Jarrett}, {Kopan}, {Kirkpatrick}, {Light}, {Marsh}, {McCallon},
  {Schneider}, {Stiening}, {Sykes}, {Weinberg}, {Wheaton}, {Wheelock}, \&
  {Zacarias}}]{Cutri2003}
{Cutri}, R.~M., {Skrutskie}, M.~F., {van Dyk}, S., {et~al.} 2003, VizieR Online
  Data Catalog, 2246

\bibitem[{{Daemgen} {et~al.}(2016){Daemgen}, {Natta}, {Scholz}, {Testi},
  {Jayawardhana}, {Greaves}, \& {Eastwood}}]{Daemgen2016}
{Daemgen}, S., {Natta}, A., {Scholz}, A., {et~al.} 2016, \aap, 594, A83

\bibitem[{{Davies} {et~al.}(2014){Davies}, {Gregory}, \&
  {Greaves}}]{Davies2014}
{Davies}, C.~L., {Gregory}, S.~G., \& {Greaves}, J.~S. 2014, \mnras, 444, 1157

\bibitem[{{de Gregorio-Monsalvo} {et~al.}(2013){de Gregorio-Monsalvo},
  {M{\'e}nard}, {Dent}, {Pinte}, {L{\'o}pez}, {Klaassen}, {Hales},
  {Cort{\'e}s}, {Rawlings}, {Tachihara}, {Testi}, {Takahashi}, {Chapillon},
  {Mathews}, {Juhasz}, {Akiyama}, {Higuchi}, {Saito}, {Nyman}, {Phillips},
  {Rod{\'o}n}, {Corder}, \& {Van Kempen}}]{deGregorio-Monsalvo2013}
{de Gregorio-Monsalvo}, I., {M{\'e}nard}, F., {Dent}, W., {et~al.} 2013, \aap,
  557, A133

\bibitem[{{Dent} {et~al.}(2013){Dent}, {Thi}, {Kamp}, {Williams}, {Menard},
  {Andrews}, {Ardila}, {Aresu}, {Augereau}, {Barrado y Navascues}, {Brittain},
  {Carmona}, {Ciardi}, {Danchi}, {Donaldson}, {Duchene}, {Eiroa}, {Fedele},
  {Grady}, {de Gregorio-Molsalvo}, {Howard}, {Hu{\'e}lamo}, {Krivov},
  {Lebreton}, {Liseau}, {Martin-Zaidi}, {Mathews}, {Meeus},
  {Mendigut{\'{\i}}a}, {Montesinos}, {Morales-Calderon}, {Mora}, {Nomura},
  {Pantin}, {Pascucci}, {Phillips}, {Pinte}, {Podio}, {Ramsay}, {Riaz},
  {Riviere-Marichalar}, {Roberge}, {Sandell}, {Solano}, {Tilling}, {Torrelles},
  {Vandenbusche}, {Vicente}, {White}, \& {Woitke}}]{Dent2013}
{Dent}, W.~R.~F., {Thi}, W.~F., {Kamp}, I., {et~al.} 2013, \pasp, 125, 477

\bibitem[{{Dullemond} \& {Dominik}(2004)}]{Dullemond2004}
{Dullemond}, C.~P., \& {Dominik}, C. 2004, \aap, 421, 1075

\bibitem[{{Feiden}(2016)}]{Feiden2016}
{Feiden}, G.~A. 2016, \aap, 593, A99

\bibitem[{{Gillon} {et~al.}(2016){Gillon}, {Jehin}, {Lederer}, {Delrez}, {de
  Wit}, {Burdanov}, {Van Grootel}, {Burgasser}, {Triaud}, {Opitom}, {Demory},
  {Sahu}, {Bardalez Gagliuffi}, {Magain}, \& {Queloz}}]{Gillon2016}
{Gillon}, M., {Jehin}, E., {Lederer}, S.~M., {et~al.} 2016, \nat, 533, 221

\bibitem[{{Gillon} {et~al.}(2017){Gillon}, {Triaud}, {Demory}, {Jehin}, {Agol},
  {Deck}, {Lederer}, {de Wit}, {Burdanov}, {Ingalls}, {Bolmont}, {Leconte},
  {Raymond}, {Selsis}, {Turbet}, {Barkaoui}, {Burgasser}, {Burleigh}, {Carey},
  {Chaushev}, {Copperwheat}, {Delrez}, {Fernandes}, {Holdsworth}, {Kotze}, {Van
  Grootel}, {Almleaky}, {Benkhaldoun}, {Magain}, \& {Queloz}}]{Gillon2017}
{Gillon}, M., {Triaud}, A.~H.~M.~J., {Demory}, B.-O., {et~al.} 2017, \nat, 542,
  456

\bibitem[{{Greenwood} {et~al.}(2017){Greenwood}, {Kamp}, {Waters}, {Woitke},
  {Thi}, {Rab}, {Aresu}, \& {Spaans}}]{Greenwood2017}
{Greenwood}, A.~J., {Kamp}, I., {Waters}, L.~B.~F.~M., {et~al.} 2017, ArXiv
  e-prints, arXiv:1702.04744

\bibitem[{{Guieu} {et~al.}(2006){Guieu}, {Dougados}, {Monin}, {Magnier}, \&
  {Mart{\'{\i}}n}}]{Guieu2006}
{Guieu}, S., {Dougados}, C., {Monin}, J.-L., {Magnier}, E., \& {Mart{\'{\i}}n},
  E.~L. 2006, \aap, 446, 485

\bibitem[{{Guieu} {et~al.}(2007){Guieu}, {Pinte}, {Monin}, {M{\'e}nard},
  {Fukagawa}, {Padgett}, {Noriega-Crespo}, {Carey}, {Rebull}, {Huard}, \&
  {Guedel}}]{Guieu2007}
{Guieu}, S., {Pinte}, C., {Monin}, J.-L., {et~al.} 2007, \aap, 465, 855

\bibitem[{{Guilloteau} {et~al.}(2011){Guilloteau}, {Dutrey}, {Pi{\'e}tu}, \&
  {Boehler}}]{Guilloteau2011}
{Guilloteau}, S., {Dutrey}, A., {Pi{\'e}tu}, V., \& {Boehler}, Y. 2011, \aap,
  529, A105

\bibitem[{{Harvey} {et~al.}(2010){Harvey}, {Jaffe}, {Allers}, \&
  {Liu}}]{Harvey2010}
{Harvey}, P.~M., {Jaffe}, D.~T., {Allers}, K., \& {Liu}, M. 2010, \apj, 720,
  1374

\bibitem[{{Harvey} {et~al.}(2012){Harvey}, {Henning}, {M{\'e}nard}, {Wolf},
  {Liu}, {Cieza}, {Evans}, {Pascucci}, {Mer{\'{\i}}n}, \& {Pinte}}]{Harvey2012}
{Harvey}, P.~M., {Henning}, T., {M{\'e}nard}, F., {et~al.} 2012, \apjl, 744, L1

\bibitem[{{Herczeg} \& {Hillenbrand}(2008)}]{Herczeg2008}
{Herczeg}, G.~J., \& {Hillenbrand}, L.~A. 2008, \apj, 681, 594

\bibitem[{{Howard} {et~al.}(2013){Howard}, {Sandell}, {Vacca}, {Duch{\^e}ne},
  {Mathews}, {Augereau}, {Barrado}, {Dent}, {Eiroa}, {Grady}, {Kamp}, {Meeus},
  {M{\'e}nard}, {Pinte}, {Podio}, {Riviere-Marichalar}, {Roberge}, {Thi},
  {Vicente}, \& {Williams}}]{Howard2013}
{Howard}, C.~D., {Sandell}, G., {Vacca}, W.~D., {et~al.} 2013, \apj, 776, 21

\bibitem[{{Isella} {et~al.}(2009){Isella}, {Carpenter}, \&
  {Sargent}}]{Isella2009}
{Isella}, A., {Carpenter}, J.~M., \& {Sargent}, A.~I. 2009, \apj, 701, 260

\bibitem[{{Kamp} {et~al.}(2011){Kamp}, {Woitke}, {Pinte}, {Tilling}, {Thi},
  {Menard}, {Duchene}, \& {Augereau}}]{Kamp2011}
{Kamp}, I., {Woitke}, P., {Pinte}, C., {et~al.} 2011, \aap, 532, A85

\bibitem[{{Keane} {et~al.}(2014){Keane}, {Pascucci}, {Espaillat}, {Woitke},
  {Andrews}, {Kamp}, {Thi}, {Meeus}, \& {Dent}}]{Keane2014}
{Keane}, J.~T., {Pascucci}, I., {Espaillat}, C., {et~al.} 2014, \apj, 787, 153

\bibitem[{{Kelly}(2007)}]{Kelly2007}
{Kelly}, B.~C. 2007, \apj, 665, 1489

\bibitem[{{Klein} {et~al.}(2003){Klein}, {Apai}, {Pascucci}, {Henning}, \&
  {Waters}}]{Klein2003}
{Klein}, R., {Apai}, D., {Pascucci}, I., {Henning}, T., \& {Waters},
  L.~B.~F.~M. 2003, \apjl, 593, L57

\bibitem[{{Liu} {et~al.}(2015){Liu}, {Joergens}, {Bayo}, {Nielbock}, \&
  {Wang}}]{Liu2015}
{Liu}, Y., {Joergens}, V., {Bayo}, A., {Nielbock}, M., \& {Wang}, H. 2015,
  \aap, 582, A22

\bibitem[{{Luhman}(2000)}]{Luhman2000}
{Luhman}, K.~L. 2000, \apj, 544, 1044

\bibitem[{{Luhman}(2008)}]{Luhman2008}
---. 2008, {Chamaeleon}, ed. B.~{Reipurth}, 169

\bibitem[{{Luhman} {et~al.}(2010){Luhman}, {Allen}, {Espaillat}, {Hartmann}, \&
  {Calvet}}]{Luhman2010}
{Luhman}, K.~L., {Allen}, P.~R., {Espaillat}, C., {Hartmann}, L., \& {Calvet},
  N. 2010, \apjs, 186, 111

\bibitem[{{Luhman} {et~al.}(2003){Luhman}, {Brice{\~n}o}, {Stauffer},
  {Hartmann}, {Barrado y Navascu{\'e}s}, \& {Caldwell}}]{Luhman2003}
{Luhman}, K.~L., {Brice{\~n}o}, C., {Stauffer}, J.~R., {et~al.} 2003, \apj,
  590, 348

\bibitem[{{Luhman} {et~al.}(2009){Luhman}, {Mamajek}, {Allen}, \&
  {Cruz}}]{Luhman2009}
{Luhman}, K.~L., {Mamajek}, E.~E., {Allen}, P.~R., \& {Cruz}, K.~L. 2009, \apj,
  703, 399

\bibitem[{{Luhman} {et~al.}(2017){Luhman}, {Mamajek}, {Shukla}, \&
  {Loutrel}}]{Luhman2017}
{Luhman}, K.~L., {Mamajek}, E.~E., {Shukla}, S.~J., \& {Loutrel}, N.~P. 2017,
  \aj, 153, 46

\bibitem[{{Luhman} {et~al.}(2007){Luhman}, {Adame}, {D'Alessio}, {Calvet},
  {McLeod}, {Bohac}, {Forrest}, {Hartmann}, {Sargent}, \&
  {Watson}}]{Luhman2007}
{Luhman}, K.~L., {Adame}, L., {D'Alessio}, P., {et~al.} 2007, \apj, 666, 1219

\bibitem[{{Luhman} {et~al.}(2008){Luhman}, {Allen}, {Allen}, {Gutermuth},
  {Hartmann}, {Mamajek}, {Megeath}, {Myers}, \& {Fazio}}]{Luhman2008a}
{Luhman}, K.~L., {Allen}, L.~E., {Allen}, P.~R., {et~al.} 2008, \apj, 675, 1375

\bibitem[{{Mart{\'{\i}}n} {et~al.}(2001){Mart{\'{\i}}n}, {Dougados}, {Magnier},
  {M{\'e}nard}, {Magazz{\`u}}, {Cuillandre}, \& {Delfosse}}]{Martin2001}
{Mart{\'{\i}}n}, E.~L., {Dougados}, C., {Magnier}, E., {et~al.} 2001, \apjl,
  561, L195

\bibitem[{{Mathis} {et~al.}(1983){Mathis}, {Mezger}, \& {Panagia}}]{Mathis1983}
{Mathis}, J.~S., {Mezger}, P.~G., \& {Panagia}, N. 1983, \aap, 128, 212

\bibitem[{{Min} {et~al.}(2009){Min}, {Dullemond}, {Dominik}, {de Koter}, \&
  {Hovenier}}]{Min2009}
{Min}, M., {Dullemond}, C.~P., {Dominik}, C., {de Koter}, A., \& {Hovenier},
  J.~W. 2009, \aap, 497, 155

\bibitem[{{Mohanty} {et~al.}(2013){Mohanty}, {Greaves}, {Mortlock}, {Pascucci},
  {Scholz}, {Thompson}, {Apai}, {Lodato}, \& {Looper}}]{Mohanty2013}
{Mohanty}, S., {Greaves}, J., {Mortlock}, D., {et~al.} 2013, \apj, 773, 168

\bibitem[{{Monin} {et~al.}(2013){Monin}, {Whelan}, {Lefloch}, {Dougados}, \&
  {Alves de Oliveira}}]{Monin2013}
{Monin}, J.-L., {Whelan}, E.~T., {Lefloch}, B., {Dougados}, C., \& {Alves de
  Oliveira}, C. 2013, \aap, 551, L1

\bibitem[{{Mordasini} {et~al.}(2012){Mordasini}, {Alibert}, {Benz}, {Klahr}, \&
  {Henning}}]{Mordasini2012}
{Mordasini}, C., {Alibert}, Y., {Benz}, W., {Klahr}, H., \& {Henning}, T. 2012,
  \aap, 541, A97

\bibitem[{{Mulders} \& {Dominik}(2012)}]{Mulders2012}
{Mulders}, G.~D., \& {Dominik}, C. 2012, \aap, 539, A9

\bibitem[{{Ott}(2010)}]{Ott2010}
{Ott}, S. 2010, in Astronomical Society of the Pacific Conference Series, Vol.
  434, Astronomical Data Analysis Software and Systems XIX, ed. Y.~{Mizumoto},
  K.-I. {Morita}, \& M.~{Ohishi}, 139

\bibitem[{{Pascucci} {et~al.}(2009){Pascucci}, {Apai}, {Luhman}, {Henning},
  {Bouwman}, {Meyer}, {Lahuis}, \& {Natta}}]{Pascucci2009}
{Pascucci}, I., {Apai}, D., {Luhman}, K., {et~al.} 2009, \apj, 696, 143

\bibitem[{{Pascucci} {et~al.}(2016){Pascucci}, {Testi}, {Herczeg}, {Long},
  {Manara}, {Hendler}, {Mulders}, {Krijt}, {Ciesla}, {Henning}, {Mohanty},
  {Drabek-Maunder}, {Apai}, {Sz{\H u}cs}, {Sacco}, \&
  {Olofsson}}]{Pascucci2016}
{Pascucci}, I., {Testi}, L., {Herczeg}, G.~J., {et~al.} 2016, \apj, 831, 125

\bibitem[{{Pilbratt} {et~al.}(2010){Pilbratt}, {Riedinger}, {Passvogel},
  {Crone}, {Doyle}, {Gageur}, {Heras}, {Jewell}, {Metcalfe}, {Ott}, \&
  {Schmidt}}]{Pilbratt2010}
{Pilbratt}, G.~L., {Riedinger}, J.~R., {Passvogel}, T., {et~al.} 2010, \aap,
  518, L1

\bibitem[{{Pinilla} {et~al.}(2013){Pinilla}, {Birnstiel}, {Benisty}, {Ricci},
  {Natta}, {Dullemond}, {Dominik}, \& {Testi}}]{Pinilla2013}
{Pinilla}, P., {Birnstiel}, T., {Benisty}, M., {et~al.} 2013, \aap, 554, A95

\bibitem[{{Poglitsch} {et~al.}(2010){Poglitsch}, {Waelkens}, {Geis},
  {Feuchtgruber}, {Vandenbussche}, {Rodriguez}, {Krause}, {Renotte}, {van
  Hoof}, {Saraceno}, {Cepa}, {Kerschbaum}, {Agn{\`e}se}, {Ali}, {Altieri},
  {Andreani}, {Augueres}, {Balog}, {Barl}, {Bauer}, {Belbachir}, {Benedettini},
  {Billot}, {Boulade}, {Bischof}, {Blommaert}, {Callut}, {Cara}, {Cerulli},
  {Cesarsky}, {Contursi}, {Creten}, {De Meester}, {Doublier}, {Doumayrou},
  {Duband}, {Exter}, {Genzel}, {Gillis}, {Gr{\"o}zinger}, {Henning},
  {Herreros}, {Huygen}, {Inguscio}, {Jakob}, {Jamar}, {Jean}, {de Jong},
  {Katterloher}, {Kiss}, {Klaas}, {Lemke}, {Lutz}, {Madden}, {Marquet},
  {Martignac}, {Mazy}, {Merken}, {Montfort}, {Morbidelli}, {M{\"u}ller},
  {Nielbock}, {Okumura}, {Orfei}, {Ottensamer}, {Pezzuto}, {Popesso},
  {Putzeys}, {Regibo}, {Reveret}, {Royer}, {Sauvage}, {Schreiber}, {Stegmaier},
  {Schmitt}, {Schubert}, {Sturm}, {Thiel}, {Tofani}, {Vavrek}, {Wetzstein},
  {Wieprecht}, \& {Wiezorrek}}]{Poglitsch2010}
{Poglitsch}, A., {Waelkens}, C., {Geis}, N., {et~al.} 2010, \aap, 518, L2

\bibitem[{{Raymond} {et~al.}(2007){Raymond}, {Scalo}, \&
  {Meadows}}]{Raymond2007}
{Raymond}, S.~N., {Scalo}, J., \& {Meadows}, V.~S. 2007, \apj, 669, 606

\bibitem[{{Rebull} {et~al.}(2010){Rebull}, {Padgett}, {McCabe}, {Hillenbrand},
  {Stapelfeldt}, {Noriega-Crespo}, {Carey}, {Brooke}, {Huard}, {Terebey},
  {Audard}, {Monin}, {Fukagawa}, {G{\"u}del}, {Knapp}, {Menard}, {Allen},
  {Angione}, {Baldovin-Saavedra}, {Bouvier}, {Briggs}, {Dougados}, {Evans},
  {Flagey}, {Guieu}, {Grosso}, {Glauser}, {Harvey}, {Hines}, {Latter},
  {Skinner}, {Strom}, {Tromp}, \& {Wolf}}]{Rebull2010}
{Rebull}, L.~M., {Padgett}, D.~L., {McCabe}, C.-E., {et~al.} 2010, \apjs, 186,
  259

\bibitem[{{Ricci} {et~al.}(2013){Ricci}, {Isella}, {Carpenter}, \&
  {Testi}}]{Ricci2013}
{Ricci}, L., {Isella}, A., {Carpenter}, J.~M., \& {Testi}, L. 2013, \apjl, 764,
  L27

\bibitem[{{Ricci} {et~al.}(2014){Ricci}, {Testi}, {Natta}, {Scholz}, {de
  Gregorio-Monsalvo}, \& {Isella}}]{Ricci2014}
{Ricci}, L., {Testi}, L., {Natta}, A., {et~al.} 2014, \apj, 791, 20

\bibitem[{{Riviere-Marichalar} {et~al.}(2016){Riviere-Marichalar},
  {Mer{\'{\i}}n}, {Kamp}, {Eiroa}, \& {Montesinos}}]{Riviere-Marichalar2016}
{Riviere-Marichalar}, P., {Mer{\'{\i}}n}, B., {Kamp}, I., {Eiroa}, C., \&
  {Montesinos}, B. 2016, \aap, 594, A59

\bibitem[{{Sawicki}(2012)}]{Sawicki2012}
{Sawicki}, M. 2012, \pasp, 124, 1208

\bibitem[{{Schaefer} {et~al.}(2009){Schaefer}, {Dutrey}, {Guilloteau}, {Simon},
  \& {White}}]{Schaefer2009}
{Schaefer}, G.~H., {Dutrey}, A., {Guilloteau}, S., {Simon}, M., \& {White},
  R.~J. 2009, \apj, 701, 698

\bibitem[{{Stassun} {et~al.}(2014){Stassun}, {Feiden}, \&
  {Torres}}]{Stassun2014}
{Stassun}, K.~G., {Feiden}, G.~A., \& {Torres}, G. 2014, \nar, 60, 1

\bibitem[{{Sz{\H u}cs} {et~al.}(2010){Sz{\H u}cs}, {Apai}, {Pascucci}, \&
  {Dullemond}}]{SzHucs2010}
{Sz{\H u}cs}, L., {Apai}, D., {Pascucci}, I., \& {Dullemond}, C.~P. 2010, \apj,
  720, 1668

\bibitem[{{Testi} {et~al.}(2016){Testi}, {Natta}, {Scholz}, {Tazzari}, {Ricci},
  \& {de Gregorio Monsalvo}}]{Testi2016}
{Testi}, L., {Natta}, A., {Scholz}, A., {et~al.} 2016, \aap, 593, A111

\bibitem[{{Torres} {et~al.}(2007){Torres}, {Loinard}, {Mioduszewski}, \&
  {Rodr{\'{\i}}guez}}]{Torres2007}
{Torres}, R.~M., {Loinard}, L., {Mioduszewski}, A.~J., \& {Rodr{\'{\i}}guez},
  L.~F. 2007, \apj, 671, 1813

\bibitem[{{Torres} {et~al.}(2009){Torres}, {Loinard}, {Mioduszewski}, \&
  {Rodr{\'{\i}}guez}}]{Torres2009}
---. 2009, \apj, 698, 242

\bibitem[{{van der Plas} {et~al.}(2016){van der Plas}, {M{\'e}nard},
  {Ward-Duong}, {Bulger}, {Harvey}, {Pinte}, {Patience}, {Hales}, \&
  {Casassus}}]{vanderPlas2016}
{van der Plas}, G., {M{\'e}nard}, F., {Ward-Duong}, K., {et~al.} 2016, \apj,
  819, 102

\bibitem[{{Woitke} {et~al.}(2009){Woitke}, {Kamp}, \& {Thi}}]{Woitke2009}
{Woitke}, P., {Kamp}, I., \& {Thi}, W.-F. 2009, \aap, 501, 383

\bibitem[{{Woitke} {et~al.}(2010){Woitke}, {Pinte}, {Tilling}, {M{\'e}nard},
  {Kamp}, {Thi}, {Duch{\^e}ne}, \& {Augereau}}]{Woitke2010}
{Woitke}, P., {Pinte}, C., {Tilling}, I., {et~al.} 2010, \mnras, 405, L26

\bibitem[{{Woitke} {et~al.}(2016){Woitke}, {Min}, {Pinte}, {Thi}, {Kamp},
  {Rab}, {Anthonioz}, {Antonellini}, {Baldovin-Saavedra}, {Carmona}, {Dominik},
  {Dionatos}, {Greaves}, {G{\"u}del}, {Ilee}, {Liebhart}, {M{\'e}nard},
  {Rigon}, {Waters}, {Aresu}, {Meijerink}, \& {Spaans}}]{Woitke2016}
{Woitke}, P., {Min}, M., {Pinte}, C., {et~al.} 2016, \aap, 586, A103

\bibitem[{{Wright} {et~al.}(2010){Wright}, {Eisenhardt}, {Mainzer}, {Ressler},
  {Cutri}, {Jarrett}, {Kirkpatrick}, {Padgett}, {McMillan}, {Skrutskie},
  {Stanford}, {Cohen}, {Walker}, {Mather}, {Leisawitz}, {Gautier}, {McLean},
  {Benford}, {Lonsdale}, {Blain}, {Mendez}, {Irace}, {Duval}, {Liu}, {Royer},
  {Heinrichsen}, {Howard}, {Shannon}, {Kendall}, {Walsh}, {Larsen}, {Cardon},
  {Schick}, {Schwalm}, {Abid}, {Fabinsky}, {Naes}, \& {Tsai}}]{Wright2010}
{Wright}, E.~L., {Eisenhardt}, P.~R.~M., {Mainzer}, A.~K., {et~al.} 2010, \aj,
  140, 1868

\end{thebibliography}

\appendix

\section{Confidence interval calculations}\label{sect:effect_of_outer_radius}

Because some of our probability distributions are asymmetric, or even
multimodal, we use a modified version of the shortest interval method for
defining our confidence intervals \citep[e.g.][]{Barlow1989,Andrae2010}.  We
restrict our confidence intervals to being continuous and move the intervals
away from the probability peak such that higher probabilities are preferred
until at least 68.27\% of the pdf area is found.

\begin{figure*}[tbh]
\makebox[\textwidth]{%
\includegraphics[width=0.55\textwidth]{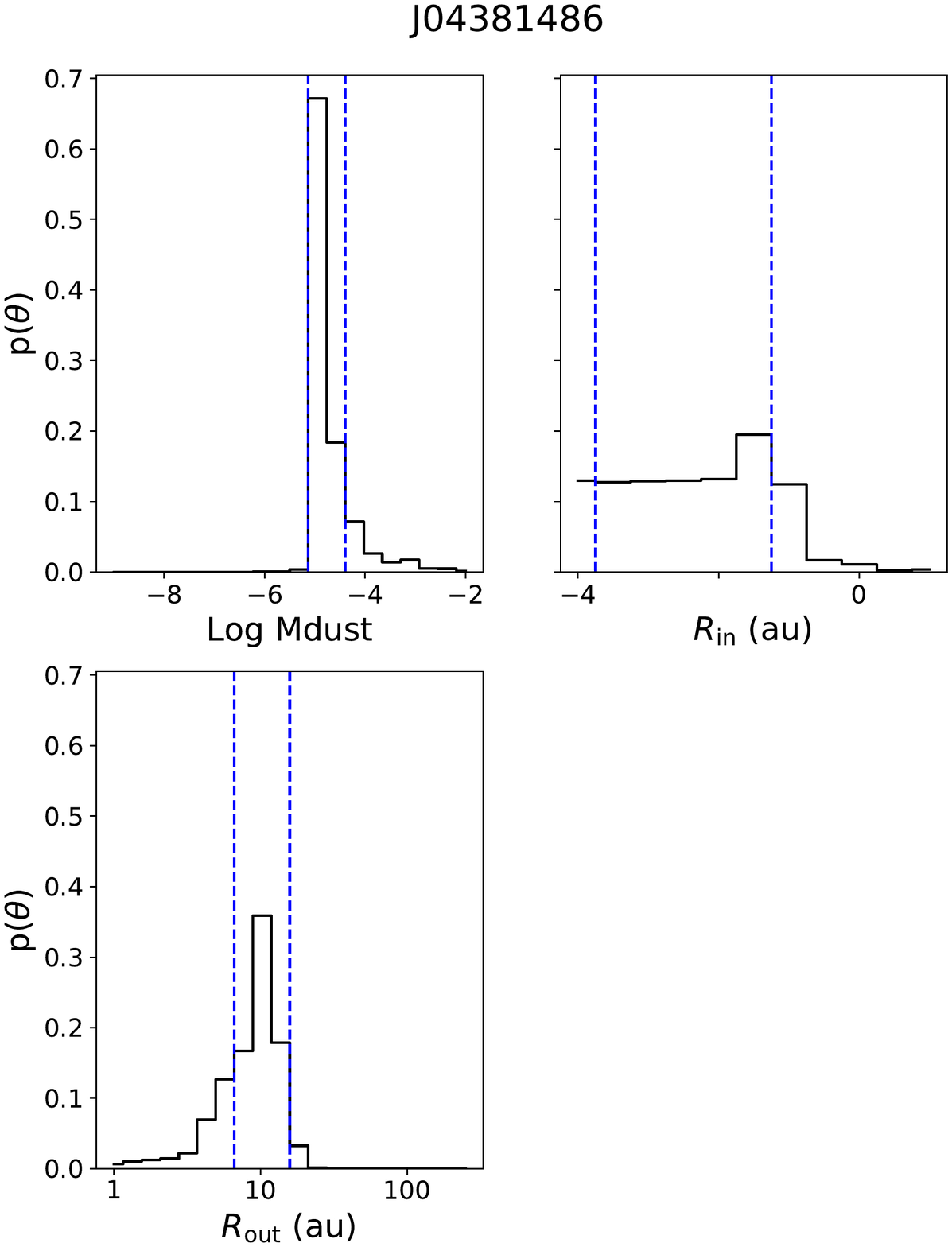}%
\hfill    
\includegraphics[width=0.55\textwidth]{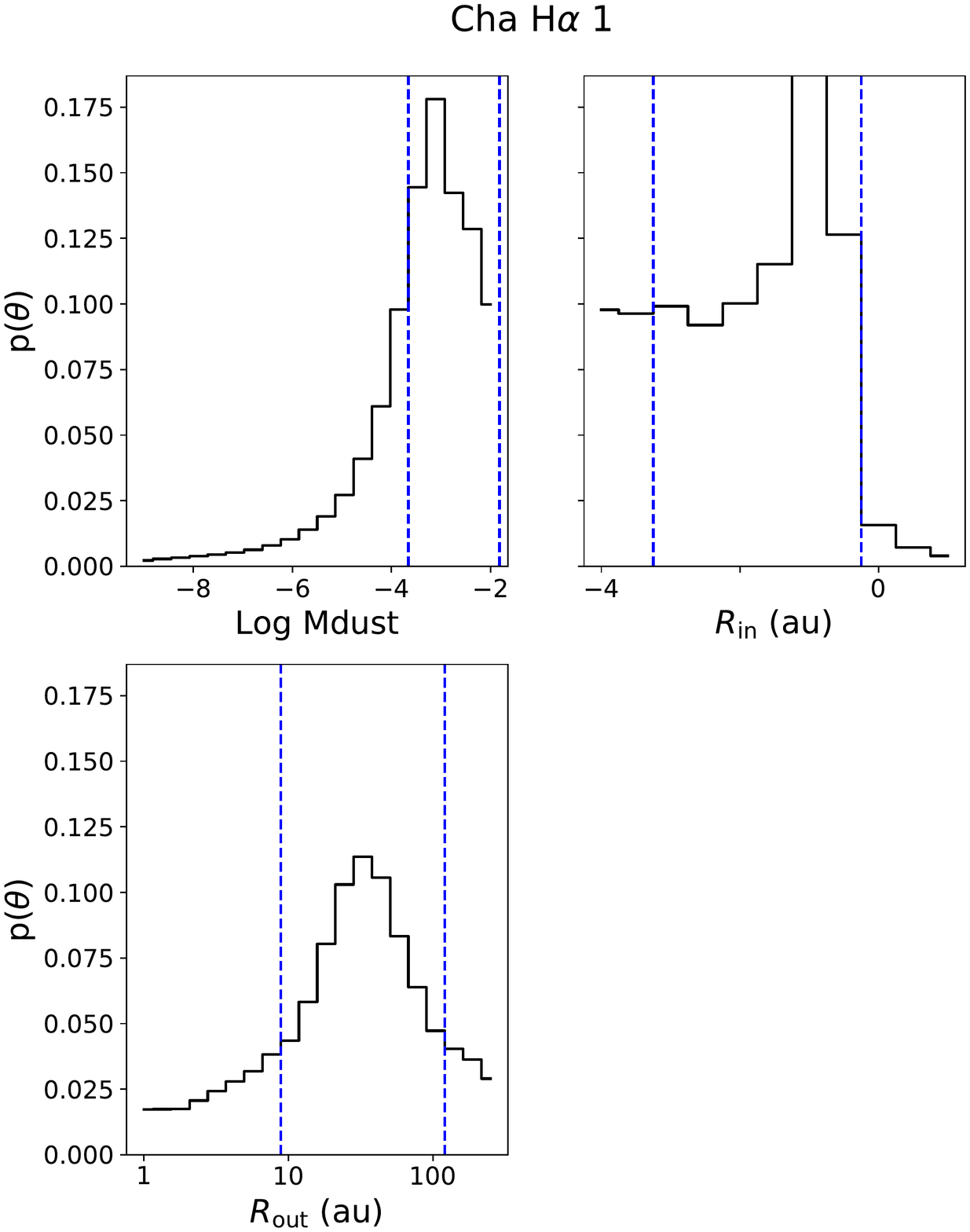}%
}
\caption{
    Bayesian probability distributions of $\Mdust$, $\Rin$ and $\Rout$ for J04381486 and Cha~H$\alpha$~1.  For these examples, we show all three of our free parameters.  The vertical dashed blue lines denote our 68.27\% confidence interval.
}
\label{figure-ci_examples}
\end{figure*}

Examples of our confidence intervals are shown for the two sources J04381486
and Cha~H$\alpha$~1 in Figure \ref{figure-ci_examples}.  These two sources
represent two extrema in the SED coverage  and illustrate how our confidence
intervals prefer areas of higher probability (e.g. the $\Rin$ panels illustrate
how the confidence intervals do not include the low probabilities to the right
of the peak probabilities), and the importance of mm photometry to constraining
the disk outer radius.  Cha~H$\alpha$~1 lacks photometry data (detections or
upper-limits) beyond 100,\micron{} and consequently the outer radius (and dust
mass) is unconstrained by our modeling.

\section{Effect of outer radius on sed}\label{sect:effect_of_outer_radius}

\begin{figure*}
    \includegraphics[width=\textwidth]{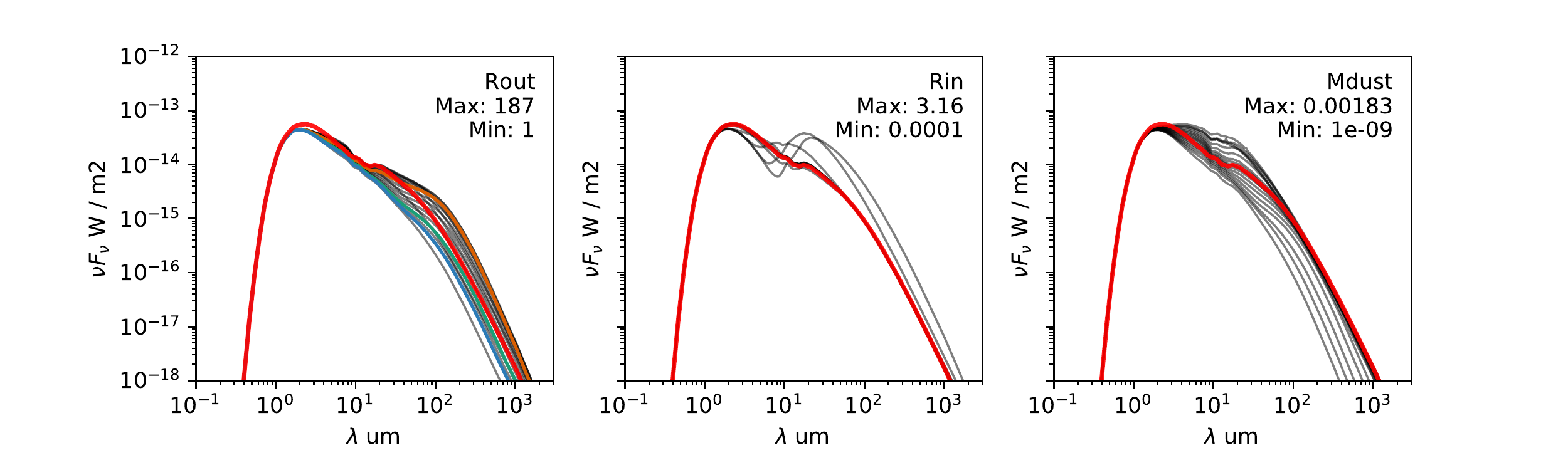}
    \caption{
        \label{figure-effect_of_params_sed-gmtau}
        In each panel we show the effect of varying the parameter ($\Rout$,
        $\Rin$ and $\Mdust$ respectively) while holding all other parameters to
        the value of the fiducial model (GM Tau; shown in red).
        Highlighted in the $\Rout{}$ plot are models with outer radii of 10~au (orange), 58~au (green) and 105~au (blue).
    }
\end{figure*}

It has been suggested that the outer radius disk parameter has a minimal effect
on the SEDs of brown dwarf disks \citep[e.g.][]{Harvey2012,Liu2015} and TT disks
\citep[e.g.][]{Woitke2016}.  Because these previous works used parameterized
scale heights and may not have extensively explored disk sizes down to very
small radii (e.g. \cite{Harvey2012} explores 50-200~au), we explore the effect
here.  Figure \ref{figure-effect_of_params_sed-gmtau} shows the dependencies of
the SED on the three disk parameters investigated in this work ($\Rout$, $\Rin$
and $\Mdust$).  At wavelengths shorter than $10,\micron$ the outer radius has
some impact on the flux density, and beyond $10,\micron$ it has significant
influence on the flux density as well as the location of the Rayleigh limit.
This example demonstrates that the outer disk radius is an important parameter
to consider in SED fitting of VLMO disks.

\end{document}